\documentclass[12pt,a4paper]{article}
\usepackage{cite,color,graphics,amsmath,epsfig,rotating}
\usepackage{siunitx}
\pdfoutput=1
   
\usepackage{latexsym}
\usepackage{amssymb}
\usepackage{slashed}
\usepackage{mathrsfs}
\usepackage{graphicx}
\usepackage{subfig}
\usepackage{psfrag}
\usepackage{url}
\DeclareMathAlphabet{\mathpzc}{OT1}{pzc}{m}{it}

\usepackage{hyperref}
\hypersetup{colorlinks=true,linktocpage,bookmarksopen,bookmarksnumbered,pdfstartview=FitH,breaklinks=true}
\ifx\pdfoutput\undefined

\usepackage[bookmarks]{hyperref}        % This is for arXiv.org
\usepackage{enumitem}
\else
\usepackage{hyperref}   % This is for pdftex
\fi

\usepackage{multicol}
\usepackage{color}
\usepackage{marginnote}
\usepackage{eurosym}

\def\micro{{\tt micrOMEGAs}}
\def\MHC{M_{H^+}}

\begin{document}
%\begin{titlepage}
\begin{center}
%\vspace*{-1cm}
%\begin{flushright}
%LAPTH-xxx/16\\
%LPSC16109
%\end{flushright}

{\bf Two dark matter candidates: the case of inert doublet and singlet scalars.}

\vspace*{1cm}\renewcommand{\thefootnote}{\fnsymbol{footnote}}

{\large  G.~B\'elanger$^{1}$, %\footnote[2]{Email: \mhref{belanger@lapth.cnrs.fr}},
A.~Mjallal$^{1}$, %\footnote[4]{Email: \mhref{mjallal@lapth.cnrs.fr}}, 
A.~Pukhov$^{2}$%\footnote[5]{Email: \mhref{pukhov@lapth.cnrs.fr}},

\renewcommand{\thefootnote}{\arabic{footnote}}

\vspace*{1cm} 
{\normalsize \it 
$^1\,$ \href{http://lapth.cnrs.fr}{LAPTh}, Univ. Grenoble Alpes, USMB,CNRS, F-74940 Annecy, France\\[2mm]
$^2\,$\href{http://theory.sinp.msu.ru}{Skobeltsyn Institute of Nuclear Physics}, Moscow State University,\\ Moscow 119992, Russia\\[2mm]
}}

\vspace{1cm}

\begin{abstract}
We consider a multi-component dark matter model where the dark sector contains a scalar doublet and a complex scalar singlet. We impose a discrete $Z_4$ symmetry to ensure  such that the lightest component of the doublet, $\tilde{A}$, and the singlet, $\tilde{S}$, are both stable. 
Interactions between the dark sectors impact significantly dark matter observables, they  allow in particular  to significantly relax the direct detection constraints on the model. To determine the  parameter space that satisfies relic density, theoretical and collider constraints as well as direct and indirect detection limits, we perform two separate scans, the first includes the full parameter space of the model while the second is dedicated to scenarios with a compressed  inert doublet spectrum. In the first case we find that the singlet is generally the dominant dark matter component while in the compressed case the doublet is more likely to be the dominant dark matter component.  In both cases we find that the two dark matter particles can have masses that ranges from around  $m_h/2$ to over the TeV scale.   We emphasize the interplay between cosmological  astrophysical  and collider constraints and show that a large fraction of the  parameter space that escapes current constraints is within the sensitivity reach of future detectors such as XENON-nT, Darwin or CTA.   Important collider signatures are mostly found in the compressed spectrum case with the possibility of probing the model with searches for heavy stable charged particles and disappearing tracks. We also show that  semi-annihilation processes such as $\tilde{S}\tilde{S}\to \tilde{A}Z$  could give the dominant signature in indirect detection searches. 
  \end{abstract} 

\end{center}
%\end{titlepage}

\section{ Introduction}

The hypothesis that a new weakly interacting particle at the electroweak scale could explain dark matter (DM) has been subjected to a host of experimental tests both in astroparticle and particle physics~\cite{Aprile:2018dbl,Fermi-LAT:2015att,Fermi-LAT:2016uux,HAWC:2019jvm,Abercrombie:2015wmb,PandaX-II:2020oim,PICO:2019vsc,Fermi-LAT:2019lyf,Acharyya:2020sbj}. Indeed, one of the most attractive feature  of these scenarios, beyond their theoretical motivation,  is the strong correlation between DM production in the early Universe and DM signatures in  direct detection (DD), indirect detection (ID) and at colliders.  These experimental searches have in recent years  severely restricted  the parameter space of typical weakly interacting massive particles (WIMP) models~\cite{Roszkowski:2017nbc,Arcadi:2017kky}. While the searches for WIMPs continue in order to cover as much as possible the large theoretical space of DM models at the same time many more avenues are being explored for DM candidates~\cite{Bertone:2018krk}. These include  extending the range of DM masses from the subGeV~\cite{Knapen:2017xzo} to the multi-TeV regions~\cite{Cirelli:2018iax} and/or  of  interaction strengths from feebly~\cite{McDonald:1993ex,Hall:2009bx,Bernal:2017kxu} to strongly interacting~\cite{Hochberg:2014dra,Hochberg:2014kqa,Lee:2015gsa,Smirnov:2020zwf}.  Another possibility is to consider a more complex dark sector that features more than one stable particle contributing to DM. 
The hypothesis of multi-component DM has been entertained for a long time, the archetype being models with a WIMP DM and another very weak or very light particle (axion, axino, gravitino)~\cite{Asaka:2000ew,Covi:2004rb,Baer:2010kd,Bae:2013hma}. Within these models, there are no significant interactions between the DM candidates and most of the studies of  the WIMP phenomenology  can be directly applied after rescaling the DM density relevant for astroparticle searches to take into account the fact that the WIMP would contribute only to a fraction of the total DM relic density. Collider signatures of these WIMPs are typically independent of the second component unless the spectrum of the dark sector is compressed and the second component which contribute to DM co-annihilation  has typical long-lived signature. Moreover specific searches for feebly and/or very light particles are being pursued, notably the searches for axion-like particles ~\cite{DiLuzio:2020wdo,Irastorza:2018dyq}, of very light DM in beam-dump experiments, or of particles that are so weakly coupled that they decay outside the detector~\cite{Agrawal:2021dbo}.  In another class of scenarios, the  interactions between the WIMP DM components can be large, thus  they can impact DM production and the relic density prediction.  For example one of the DM could even be decoupled from the SM and  be thermalised only  through its interactions with the second dark sector, a process dubbed assisted freeze-out ~\cite{Belanger:2011ww}. More generally interactions between two dark sectors would contribute to DM formation even when both interact with the SM ~\cite{Belanger:2012vp,Belanger:2020hyh}. Processes such as semi-annihilation~\cite{Hambye:2008bq,Hambye:2009fg,DEramo:2010keq} where two DM annihilate into a DM and a SM particle, or processes called dark annihilation or DM conversion where a pair of DM particles annihilate into another pair of DM particles ~\cite{Arcadi:2016kmk} as well as the decay of one  DM into another component can  all play an important role in the early Universe. Examples of simple extensions of the SM that have these features include  models with two scalars~\cite{Drozd:2011aa,Drozd:2011pex,Belanger:2012vp,Biswas:2013nn,Bhattacharya:2016ysw,Bhattacharya:2017fid,Chakrabarty:2021kmr},  two fermions~\cite{Belanger:2011ww,Cirelli:2010nh,Heeck:2012bz,Kajiyama:2013rla,Gu:2013iy,Bernal:2018aon} or two vectors~\cite{Gross:2015cwa,Karam:2016rsz} as DM candidates, models with a combination of  scalar, fermion and vector DM~\cite{Daikoku:2011mq,Bhattacharya:2013hva,Bian:2013wna,Esch:2014jpa,DiFranzo:2016uzc,Ahmed:2017dbb,Chakraborti:2018lso,Poulin:2018kap} as well as other multi-component  models~\cite{Profumo:2009tb,Zurek:2008qg,Elahi:2019jeo,Khalil:2020syr}. In such models the DM components can be  stabilized by a discrete symmetry such as $Z_2\times Z_2$~\cite{Cheon:2008sym,Borah:2019aeq,Biswas:2019ygr,Bhattacharya:2019fgs} or $Z_N$ with $N>4$~\cite{Batell:2010bp,Ivanov:2012hc,Belanger:2014bga,Yaguna:2019cvp,Aranda:2019vda,Belanger:2020hyh}
The modifications to DM formation not only impacts the value of the relic density but also of various observables such as DM direct detection through DM scattering in a large underground detector, and/or specific indirect detection signals from DM annihihilation in the galaxy, as will be detailed below. Moreover structure formation and other cosmological simulations were performed within multi-component DM scenarios ~\cite{Malekjani:2007pv,Semenov:2013jhd,Medvedev:2013vsa}. One of the advantage of 
multi-component DM models is that they can explain observed cosmological and astrophysical anomalies  that require DM with different mass scales.  Typically the self-interacting DM that can fit  the observations  on small cosmological scales is very light~\cite{Rocha:2012jg,Weinberg:2013aya,Spergel:1999mh}  while DM that can explain the gamma-ray excess in the Fermi-LAT data is rather at the electroweak scale~\cite{Fermi-LAT:2009ihh,Calore:2014nla}.

In this paper we consider a model which contains both  a doublet and a singlet of scalars
 in addition to the SM particles, the inert doublet and singlet model (IDSM). We impose a discrete $Z_4$ symmetry and require different charges for the doublet and the singlet, this guarantees the stability of the singlet and of the lightest neutral component of the doublet if its decay to two singlets is kinematically forbidden.  Moreover, in the specific case of a very compressed spectrum there can also be a third DM candidate when the second neutral component of the doublet is almost stable. This model, in the limit where the singlet and doublet decouple, incorporates the cases of the much-studied inert doublet~\cite{Deshpande:1977rw,Barbieri:2006dq,LopezHonorez:2006gr} and scalar singlet~\cite{Urbano:2014hda,Feng:2014vea,Goudelis:2009zz,Yaguna:2008hd,Han:2016gyy,Mambrini:2011ik} dark matter models.  
Having a second dark matter component contributes to relaxing several of the constraints on the sub-models even when imposing that the total relic density completely matches the value determined precisely by PLANCK~\cite{Planck:2015fie,Planck:2018vyg}.   In the scalar singlet model, the coupling that is necessary to provide efficient annihilation of DM in the early universe is the one that is responsible for DM interactions with nuclei, the two requirements are compatible only for a narrow window when the DM mass is close to half the Higgs mass or for  heavy DM above the TeV scale~\cite{Cheung:2012xb,Cline:2013gha,Feng:2014vea,Athron:2017kgt,Athron:2018ipf,Athron:2018hpc}. Interactions between the two dark sectors adding processes such as DM conversion and semi-annihilation~\cite{Hambye:2008bq,Hambye:2009fg,DEramo:2010keq} means that a reduced coupling of the singlet can be sufficient to prevent over-production of DM in the early Universe,  thus loosening the direct detection constraints as will be detailed later.

The Inert doublet model is also subject to strong constraints if the DM mass is below the W mass. Here again efficient enough annihilation in the early universe requires either a large coupling to the Higgs or co-annihilation with the heavier members of the doublet,  the former is constrained by direct detection and by invisible Higgs decays while the latter by collider constraints from LEP on charged Higgs~\cite{Belyaev:2016lok}. On the other hand as soon as the WW channel opens up annihilation becomes very efficient, typically the relic density lies below the measured value unless DM is at the TeV scale~\cite{Belyaev:2016lok,Goudelis:2013uca,Arhrib:2013ela,Diaz:2015pyv}. The IDM therefore provides typically only a sub-component of the total DM. In addition to DD, the model is also constrained by indirect detection~\cite{Eiteneuer:2017hoh,Modak:2015uda,Queiroz:2015utg,Garcia-Cely:2015khw} as well as by LHC searches for monojets, multi-leptons and or invisible  Higgs ~\cite{Krawczyk:2013jta,Belanger:2015kga,Miao:2010rg,Hashemi:2016wup,Gustafsson:2012aj,Poulose:2016lvz,Ilnicka:2015jba,Blinov:2015qva}. The presence of another DM component will not only provide a complete picture for DM but also  the interactions with another DM component will alter the phenomenology of the IDM.

Our goal is first to make an up-todate analysis of the constraints on the Z4IDSM model including   theoretical constraints such as perturbativity and vacuum stability as well as experimental constraints at colliders (scalar searches at LEP,  electroweak precision, Higgs properties and searches for new particles at LHC) and in astroparticle searches (direct and indirect detection). At the LHC the most powerful searches apply in the case of light DM (below $m_h/2$) and in the case of a  compressed doublet spectrum when searches for disappearing tracks and heavy stable charged particles are relevant. Direct detection searches, in particular XENON1T, contribute to strongly constrain the model. Current Indirect detection searches with photons have more moderate  impact because the  dominant DM component is typically rather heavy while they constrain mainly  the region where DM is at the electroweak scale. Indirect searches with antiprotons allow  to probe the model although these searches are dependent both on the DM profile and the cosmic ray propagation parameters.
After having ascertain the current viability of the model we then discuss the potential of future searches to further probe the model. Among these we include prospects for direct detection searches with XENON-nT~\cite{XENON:2015gkh} and DARWIN~\cite{DARWIN:2016hyl} as well as  indirect detection with CTA~\cite{Acharyya:2020sbj}. We also comment on the potential of  improved searches for invisible Higgs  and for  monojets at colliders   although these are of limited interest after taking into account all astrophysical constraints.

The paper is organised as follows. The model is described in Section 2  and dark matter observables in section 3. Section 4 reviews various constraints on the model  and the results of the parameter scan are presented in Section 5 both for the generic case and  for the compressed doublet scenario.  Section 6 contains our conclusions.

\section{The model}
\label{model}

The model features the gauge symmetry and the fermionic content of the SM while the scalar sector contains in addition to  the SM scalar doublet Higgs $H_{1}$, an inert doublet $H_{2}$ and a complex singlet $S$. We assume that all particles of the SM, including   $H_1$, are  invariant under a discrete  $Z_4$ symmetry while the extra doublet and singlet have different charges. This choice provides the simplest way to have two DM candidates, the singlet and the  lightest component of the doublet. 
 This leaves two  possibilities for the charge assignment, $X$, where the scalar  fields transform as $\phi \to \exp^{(i X \pi/2)} \phi$, namely 
 $X_S=2$; $X_{H_2}=1,3$  or   $X_S=1,3$; $X_{H_2}=2$, in both cases $X_{H_1}=0$. We will focus on the latter choice as it offers more possibilities for DM formation, namely
it  allows for semi-annihilations where two DM particles  annihilate into a DM and a SM particle~\cite{Hambye:2008bq,Hambye:2009fg,DEramo:2010keq}.  
With this choice of charge assignment the first dark sector contains the complex singlet \footnote{Note that a mass term that split the real and imaginary components of the singlet is not allowed by the $Z_4$ symmetry for this charge assignment. } while the 4 particles of the second doublet belong to the second dark sector.  
The scalar potential reads 
\begin{eqnarray}
V_{Z4}&=&\lambda_1\left(\left|H_1 \right|^2-\frac{v^2}{2}  \right)^2 + \mu_2^2\left|H_2^2 \right| +\lambda_2\left| H_2\right|^4 + \mu_S^2 \left|S \right| ^2 + \lambda_S \left| S\right|^4 +\frac{\lambda^\prime_S}{2}\left(S^4+S^{\dagger 4} \right) \nonumber\\
&&+ \lambda_{S1} \left|S \right|^2 \left| H_1\right| ^2 + \lambda_{S2} \left|S \right| ^2 \left|H_2 \right| ^2 \nonumber\\
&&+ \lambda_3 \left|H_1 \right| ^2 \left| H_2\right| ^2 + \lambda_4 \left(H_1^\dagger H_2 \right)\left(H_2^\dagger H_1 \right)     
+ \frac{\lambda_5}{2}\left[\left(H_1^\dagger H_2 \right)^2 + \left(H_2^\dagger H_1 \right)^2\right]  \nonumber\\&&+\frac{\lambda_{S12}}{2} \left( S^2H_1^\dagger H_2 + S^{\dagger 2} H_2^\dagger H_1\right) + \frac{\lambda_{S21}}{2}\left(S^2H_2^\dagger H_1 + S^{\dagger 2} H_1^\dagger H_2 \right)
\label{eq:VZ4}
\end{eqnarray}

The scalar sector of the model has 13 independent parameters. We will use the five masses of the scalars, $M_h$ for the SM Higgs which we fix at 125GeV, $M_H,M_A,M_{H^\pm}$ for the two neutral and charged doublet and $M_S$ for the singlet, as well as eight couplings 

\begin{equation}
 \lambda_2,\lambda_3,\lambda_S,\lambda'_S,\lambda_{S1},\lambda_{S2},\lambda_{S12},\lambda_{S21}
 \label{eq:free}
\end{equation}
The remaining parameters of the potential can be simply related to these through,
\begin{eqnarray}
\lambda_1=\frac{M_h^2}{2v^2} \;;\;\;\;   \mu_2^2= M^2_{H^\pm}\;;\;\;\; \mu_S^2=M_S^2-\lambda_{S1}\frac{v^2}{2} \nonumber\\
\lambda_4 = \frac{M_H^2+M_A^2-2M_{H^\pm}^2}{v^2} \;;\;\;\; \lambda_5 = \frac{M_H^2-M_A^2}{v^2}
\end{eqnarray}
where $v$ is the vacuum expectation value of the SM doublet $H_1$. For the doublet the mass parameters can also be replaced with the mass difference with DM,
\begin{equation}
\Delta^+=M_{H+} -M_A \;,\;\;\;\;  \Delta^0=M_{H} -M_A
\end{equation}
Among the couplings in Eq.~\ref{eq:free}, the quartic self-couplings $\lambda_S,\lambda'_S$ for the singlet and $\lambda_2$ for the doublet  will play little role in DM observables.
In the inert doublet model it is customary to choose as independent parameter the coupling $\lambda_{Hh}=\lambda_3+\lambda_4+\lambda_5$ which determines the $h\tilde H^0 \tilde H^0$ coupling, $-\frac{2M_W}{g} \lambda_{Hh}$, or $\lambda_{Ah}=\lambda_3+\lambda_4-\lambda_5$ which sets the $h\tilde A \tilde A$ coupling given by $-\frac{2M_W}{g} \lambda_{Ah}$. These are simply related to our independent parameters,
\begin{equation}
\lambda_{Hh}=\lambda_3-\frac{2}{v^2} \left(M^2_{H^\pm}-M^2_H  \right) \;;\;\;\; \lambda_{Ah}=\lambda_3-\frac{2}{v^2} \left(M^2_{H^\pm}-M^2_A  \right) 
\end{equation}
Other trilinear couplings that will be relevant for the following analyses are the coupling $h\tilde{H}^+\tilde{H}^-$, $-\frac{2M_W}{g} \lambda_3$ as well as the couplings involved in semi-annihilation,
the trilinear coupling $\tilde{S}\tilde{S}\tilde{A}$,  $i M_W/g (\lambda_{S21} -  \lambda_{S12})$, and  the coupling  $\tilde{S}\tilde{S}\tilde{H}$,
 $ M_W/g (\lambda_{S21} + \lambda_{S12})$.
Note that in order to have two stable DM candidates, the decay of the neutral inert doublet into the singlet must be kinematically forbidden,  that is $M_{A/H}< 2M_S$.  When this condition is violated only $\tilde{S}$ is the DM and the model effectively reduces to the scalar singlet model as will be discussed in Section ~\ref{sec:results}. In this model, the DM phenomenology is similar whether the scalar or the pseudoscalar component of the doublet is the DM candidate, in the following we will choose the pseudoscalar as DM. 

We have also included in the model  effective vertices for $\tilde{H}^\pm\tilde{A}\pi^\mp$ and $\tilde{H}^\pm\tilde{H}\pi^\mp$ interactions. These  vertices  become relevant when the mass splitting between the charged and neutral Higgses is below a few hundred MeV's~\cite{Chen:1996ap}. In this case the dominant decay mode is $\tilde{H}^\pm\rightarrow \pi^\pm \tilde{A}$, which has a much larger decay width then the one computed for decay into quarks $\tilde{H}^+\rightarrow  u\bar{d} A$. To this end, following Ref.~\cite{Belyaev:2016lok,Belyaev:2020wok}, we introduce the non perturbative $W-\pi$ mixing through the Lagrangain
\begin{equation}
{\cal L}= \frac{g f_\pi}{2\sqrt{2}} W_\mu^+\partial^\mu \pi^-
\end{equation}
such that the $\tilde{H}^+\tilde{A}\pi^-$ vertex   is defined as 
\begin{equation}
\frac{g^2 f_\pi}{4\sqrt{2} M_W^2} (p_{H^+} -p_{A}) \cdot p_\pi 
\end{equation}
where $f_\pi=130$~MeV. For decays that take place on-mass shell the vertex reduces to $g^2 f_\pi /4\sqrt{2} (M_{H+}^2-M_A^2)/M_W^2$. 
This effective vertex is taken into account only if $\Delta^+$ lies in the range $140 MeV\leq \Delta^+ \leq 500 MeV$. For smaller mass splittings the charged Higgs decay into leptons and a neutral Higgs,  for larger mass splittings  the decay of the charged Higgs is prompt and the  decay width into quarks gives an accurate enough result as was shown in ~\cite{Goodsell:2020lpx}.

For completeness we mention that there is an alternative charge assignment corresponding to $X_S=2$ and $X_H=1$, however in this case semi-annihilation processes are not allowed. The phenomenology of this model will be discussed in a separate publication.
Finally we stress that we treat the model at tree-level and we assume that all parameters are input at the electroweak scale. We have estimated the loop-corrections to the masses of the scalar sector and found that small mass splittings will be induced even for a completely degenerate tree-level spectrum, typically the mass split with the DM lies  below $\approx 500$MeV (50MeV) for the charged (neutral) component~\cite{Ali_these}. For self-consistency with the relic density calculation which is performed at tree-level we therefore chose  a complete tree-level treatment, keeping in mind  that small mass splittings are justified as arising from one-loop corrections.
For a complete discussion of the impact of one-loop corrections on the relic density in the inert doublet model see Ref.~\cite{Banerjee:2019luv,Banerjee:2021anv,Banerjee:2021xdp,Banerjee:2021hal,Banerjee:2021oxc}. 
The one-loop corrections to the 3-body decay width of the charged Higgs into DM and  light leptons were computed for a few typical points and found to be around the percent level~\cite{Shankha} thus justifying a tree-level treatment.

\section{Dark matter relic density}
 \label{sec:relic}
 
The relic density of DM provides one of  the most important constraints on the model after we impose the condition that the two DM candidates must account for all  of the DM.   Before discussing in general terms DM production in the early Universe within the $Z_4$ model,    we first briefly review the main characteristics of DM in the IDM and in the scalar singlet model, which in the limit where the two sector decouple form the two components of DM  in the Z4IDSM model.

In the scalar singlet model, all annihilation channels depend on $\lambda_{S1}$ the coupling between the scalar singlet and the SM Higgs, thus   the relic density basically fixes this parameter to $\lambda_{S1}\approx 0.15 (M_S/1000 GeV)$ except in  the region $M_S\approx m_h/2$ where it can be much smaller. However the same coupling also enters the direct detection cross-section.  Typically  a coupling large enough to provide sufficient DM annihilation in the early Universe  leads to a direct detection cross-section in conflict with XENON1T results unless the DM is  above the TeV scale, a narrow region around $m_S\approx m_h/2$ that satisfies DD constraints also remains ~\cite{Athron:2018ipf,Athron:2017kgt,Feng:2014vea,Cline:2013gha,Cheung:2012xb}.

In the IDM, the relic density is driven mostly by gauge couplings and by the DM coupling to the Higgs, $\lambda_{Hh}$ or $\lambda_{Ah}$ for the scalar or pseudoscalar DM respectively.  
Typically DM is under abundant when annihilation into W pairs are  kinematically accessible. An important exception is found  when the spectrum is compressed and $M_{DM} \geq 500$~GeV, in this case the annihilation into longitudinal gauge bosons can be suppressed, thus allowing for a relic density, $\Omega h^2 \approx 0.1$~\cite{Goudelis:2013uca}. 
Below the W threshold, the inert Higgs can account for all of the DM  if its coupling to the Higgs is large - this is however constrained from direct detection.  In this region, underabundant DM  is also found if the mass of the inert Higgs is such that  $M_A\approx m_h/2$ and the DM  annihilation cross-section is enhanced by a resonance or 
for a compressed spectrum which  benefit from many coannihilation channels.  Note that the spin-independent (SI)  direct detection cross-section also depends  on $\lambda_{Hh}$ ($\lambda_{Ah}$)  so that the largest values of this coupling are excluded even when the inert doublet is only a subcomponent of DM.
 
 A model with an inert doublet and singlet helps relaxes these constraints, in particular the additional processes participating in the DM singlet annihilation (DM conversion or semi-annihilation) means that lower values of   $\lambda_{S1}$ are allowed, thus allowing to avoid strong direct detection constraints on the singlet DM component. Moreover the same process can also reduce the fraction of the doublet component hence helping evade direct detection constraints.
  In the next subsection we illustrate the impact on the relic density of the interactions between the two dark sectors.

\subsection{Interactions between dark sectors}
	
 In the following we split the particle content in different sectors according to their $Z_4$ charge.  The dark sector 1 contains the singlet, the dark sector 2 contains the doublet while all SM particles are in sector  $0$.  The equations for the yields of the first DM, here the singlet,  and the second DM, the doublet,  read
	\begin{eqnarray}
	    3H\frac{dY_1}{ds}&&=\langle \sigma_v^{1100} \rangle \left(Y_1^2-\bar{Y}_1^2\right) +\langle \sigma_v^{1120} \rangle\left(Y_1^2-Y_2\frac{\bar{Y}_1^2}{\bar{Y}_2}\right) + \langle\sigma_v^{1122} \rangle\left(Y_1^2-Y_2^2\frac{\bar{Y}_1^2}{\bar{Y}_2^2}\right)\\
	    3H\frac{dY_2}{ds}&&=\langle\sigma_v^{2200}\rangle \left(Y_2^2-\bar{Y}_2^2\right)-\frac{1}{2} \langle\sigma_v^{1120} \rangle\left(Y_1^2-Y_2\frac{\bar{Y}_1^2}{\bar{Y}_2}\right) + \langle \sigma_v^{1210} \rangle Y_1\left(Y_2-\bar{Y}_2\right) \nonumber\\
	    &&+ \langle\sigma_v^{2211}\rangle \left(Y_2^2- Y_1^2\frac{\bar{Y}_2^2}{\bar{Y}_1^2}\right) 
	\end{eqnarray}
where $Y_{1,2}$ are the abundances of each DM, $\bar{Y}_{1,2}$, the corresponding equilibrium abundances, $s$ is the entropy density, $H$ the Hubble parameter and $\langle \sigma_v^{ijkl}\rangle$ the thermally averaged cross-sections for all processes of the type $i,j\rightarrow k,l$ involving particles in sectors  0,1 and 2. In this equation there is an implicit summation over all particles that are involved in a given subprocess, e.g. $\sigma_v^{2200}$ includes all processes involving the annihilation of pairs of $\tilde{H^+},\tilde{H},\tilde{A}$  into pairs of SM particles.
Note that   the thermally averaged cross-section $\sigma_v^{ijkl}$  satisfies the balance
equation
\begin{equation}
  (1-\frac{1}{2}\delta_{ij})  \bar{Y}_i\bar{Y}_j\langle\sigma_v^{ijkl}\rangle=(1-\frac{1}{2}\delta_{kl})\bar{Y}_k\bar{Y}_l\langle\sigma_v^{klij}\rangle.
\end{equation}

The interactions between the two dark sectors leading to DM conversion  ($\sigma_v^{1122}, \sigma_v^{2211}$) and the presence of semi-annihilation terms, $\sigma_v^{1120}$,   can strongly affect the relic density of each DM candidate while  semi-annihilation of the type  $\sigma_v^{1210}$, will only impact the second DM. Processes which contribute to DM conversion, e.g. $\tilde{S}\tilde{S^\dagger}\rightarrow \tilde{H_i}\tilde{H_j}$, are driven either  by quartic interactions which depend on the coupling $\lambda_{S2}$ or by annihilation through the SM Higgs which is proportional to $\lambda_{Ah} \lambda_{S1}$.To illustrate 
the effect of dark matter conversion,  we neglect semi-annihilation, for this  we choose a benchmark where  $\lambda_{S12}=\lambda_{S21}=0$. Moreover   we fix  $M_S=250$~GeV,  $M_A=120$ GeV, $M_{H^+}=M_H=125$ GeV, $\lambda_{S1}=10^{-3}$, $\lambda_3=8\times 10^{-3}$ and thus ${\lambda}_{Ah}=-3.2\times 10^{-2}$. For such choices of couplings processes which contribute to DM conversion  are driven by the quartic coupling $\lambda_{S2}$ since the annihilation through the SM Higgs proportional to $\lambda_{Ah} \lambda_{S1}$ is suppressed. 

\begin{figure}[h]
	\includegraphics[scale=0.45]{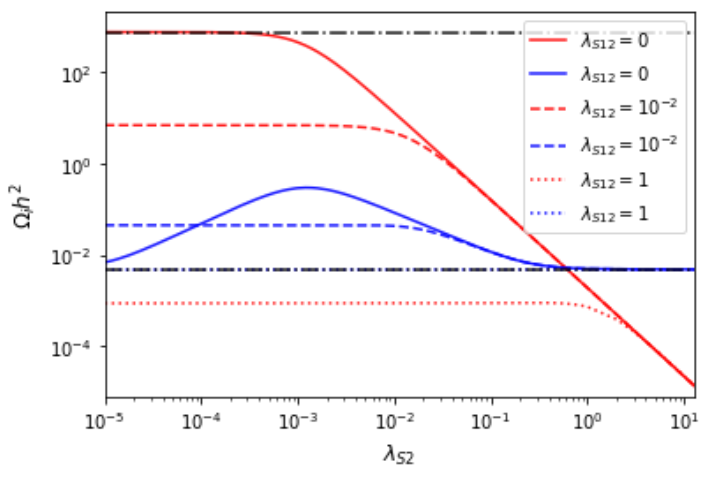}
	\includegraphics[scale=0.45]{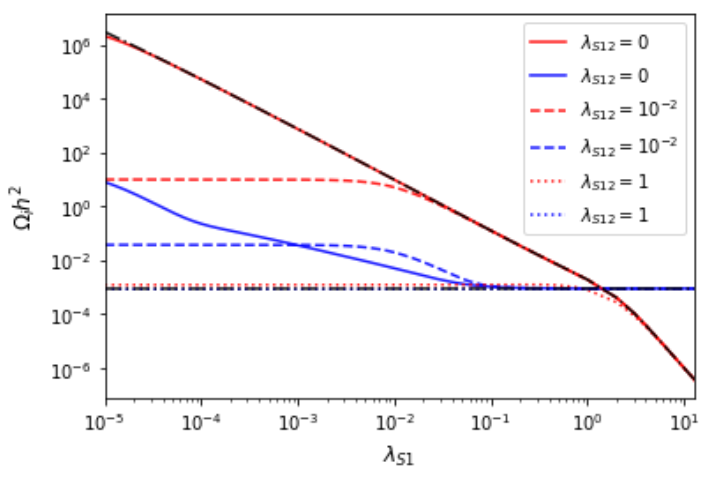}
		\centering
	\caption{Dependence of $\Omega_1 h^2$ and $\Omega_2 h^2$ on $\lambda_{S2}$ (left) and $\lambda_{S1}$  (right) for  $\lambda_{S12}=-\lambda_{S21}=0,0.01,1$, $M_A=120$ GeV, 
	$M_{S}=250$~GeV and  $\lambda_3=8\times 10^{-3}$. Other parameters are set to (left) 
	 $M_{H^+}=M_{H}=125$ GeV,  $\lambda_{S1}=10^{-3}$, and  (right)  $M_{H^+}=M_{H}=200$ GeV, $\lambda_{S2}=10^{-5}$.}
	 	\label{fig:lambdaS2}
\end{figure}

Because the coupling of the singlet is small,  $\lambda_{S1}<<1$, its relic density is very large if in addition $\lambda_{S2}<<1$, see Fig.~\ref{fig:lambdaS2} (left panel). Since $\tilde{S}$ freezes-out when $\tilde{A}$ is still in thermal equilibrium, $11\rightarrow22$	processes act as  additional annihilation channels for $\tilde{S}$ thus we observe a sharp decrease in $\Omega_1$ when $\sigma_v^{1122}$ becomes larger than $\sigma_v^{1100}$, this occurs for  $\lambda_{S2}>10^{-3}$. 
The behaviour of $\Omega_2$ is different. First note that because of  gauge interactions with the SM sector, $\Omega_2$ is typically quite small. Turning on the interactions between the dark sectors  leads to an increase of $\Omega_2$ with $\lambda_{S2}$ due to $11\rightarrow 22$ processses. Increasing  $\lambda_{S2}$ further   this channel progressively becomes less efficient because the abundance  $Y_1$ decreases sharply, the turning point is around $\lambda_{S2}=10^{-3}$. For larger values of $\lambda_{S2}$,   $\Omega_2$ is mostly set by the interactions  of the second dark sector  with the SM.   This is the case even when $\lambda_{S2}$ becomes ${\cal O}(1)$ since the kinematic suppression of the channels $22\rightarrow 11$ prevents a further drop in $\Omega_2$. 

 The impact of $\lambda_{S12}$ on $\Omega_1$ and $\Omega_2$ is similar to that of $\lambda_{S2}$ in the sense that semi-annihilation processes lead to a decrease of $\Omega_1$, see Fig.~\ref{fig:lambdaS2}. Here we choose $\lambda_{S12}=-\lambda_{S21}$ to maximise the coupling to the pseudoscalar DM, $\tilde{S}\tilde{S}\tilde{A}$. For moderate values of $\lambda_{S12}=-\lambda_{S21}=0.01$, $\Omega_2$ increases due to semi-annihilation processes such as $\tilde{S}\tilde{S}\rightarrow \tilde{A}h$ or  $\tilde{S}h\rightarrow  \tilde{S}^\dagger \tilde{A}$ as can be seen in Fig.~\ref{fig:lambdaS2}  in the limit of small $\lambda_{S2}$.  For larger  values of $\lambda_{S12}=1$ semi-annihilation does not impact $\Omega_2$ 
because the abundance $Y_1$ is small, for the same reason $\Omega_2$ is not sensitive to DM conversion and is thus  mostly independent of $\lambda_{S2}$. Moreover $\Omega_1$ is mostly dominated by  standard annihilation processes $11\rightarrow 00$ and semi-annihilation and decreases because of  DM conversion only for large values of $\lambda_{S2}$.
 
 For the  second  benchmark we increase the mass splitting between the doublet components fixing  $M_{H^+}=M_H=200$ GeV, this increases significantly 
 ${\lambda}_{Ah}=-0.84$,  we also fix $\lambda_{S2}=10^{-5}$. Since $\lambda_{S2}$ is small,  DM conversion processes proceed through the s-channel SM Higgs exchange, e.g. $\tilde{S}\tilde{S^\dagger}\rightarrow h\rightarrow \tilde{H_i}\tilde{H_j}$ and play an important role in determining the abundance of the doublet. Since $\tilde S$ is heavier than $\tilde A$, DM conversion has little impact on the singlet component.  For example for $\lambda_{S1}=10^{-5}$,  $\Omega_2 h^2=7.7$ when all processes are included, however if we switch off the DM conversion processes $\Omega_2 h^2$ drops to  $1.1\times 10^{-3}$. The impact of DM conversion is not so drastic for larger values of $\lambda_{S1}$ because  it depends on $Y_1$ which is suppressed at larger $\lambda_{S1}$. In fact  when $Y_1$ becomes small, that is for large $\lambda_{S1}$ we expect  little dependence of $\Omega_2$ on $\lambda_{S1}$, as seen in Fig.~\ref{fig:lambdaS2} (right panel).
In the limit where semi-annihilation can be neglected ($\lambda_{S12}=0$),  $\sigma v^{1100}$ gives the main  contribution to  $\Omega_1$, hence the decrease when $\lambda_{S1}$ increases,    at the same time,  $\Omega_2$ decreases because the effect of DM conversion becomes less important as $Y_1$ becomes smaller. 
Note that when $\lambda_{S1}$ becomes very large, the annihilation channel $\tilde{S}\tilde{S}^\dagger\to hh$ becomes dominant as the t-channel diagram  grows as $\lambda_{S1}^4$.
As in the previous example,  semi-annihilation terms decrease $\Omega_1$ and have the largest impact at small values of $\lambda_{S1}$ where $Y_1$ is large. Moreover semi-annihilation typically also lead to a decrease of $\Omega_2$.  However,  as explained above, semi-annihilation processes such as $\tilde{S}\tilde{S}\to \tilde{A}h$ or $\tilde{S} h\to \tilde{S}^\dagger \tilde{A}$ can lead to an increase in $\Omega_2$  if $Y_1$ is sizeable,  this occurs  for moderate values of $\lambda_{S12}=-\lambda_{S21}=0.01$.

We will also consider the case where the two neutral components of the neutral doublet are nearly degenerate. In this case  we solve  three components equations. As usual the singlet belongs to the first dark sector, $\tilde{A}$ and $\tilde{H}^\pm$ to the second dark sector and $\tilde{H}$  to a third dark sector. The evolution equations for three components are easily generalised to,

\begin{eqnarray}
\label{boltzmann1}
\nonumber
3H\frac{dY_1}{ds}&=& \langle\sigma_v^{1100}\rangle  \left(Y_1^2-\bar{Y}_1^2 \right) 
                  +  \sum_{k=2,3} \left[   \langle\sigma_v^{11k0}\rangle\left( Y_1^2 - Y_k \frac{\bar{Y}_1^2}{\bar{Y}_k} \right)
                                         + \langle\sigma_v^{11kk}\rangle\left( Y_1^2- Y_k^2  \frac{\bar{Y}_1^2}{\bar{Y}_k^2}\right) 
                                 \right] \nonumber\\
               &+&  \langle\sigma_v^{1123}\rangle \left( Y_1^2 - Y_2 Y_3 \frac{\bar{Y}_1^2}{\bar{Y}_2\bar{Y}_3} \right) 
\\
\nonumber
3H\frac{dY_k}{ds}&=&  \langle\sigma_v^{kk00}\rangle  \left(Y_k^2-\bar{Y}_k^2 \right) 
                   +  \langle\sigma_v^{kk11}\rangle\left(Y_k^2- Y_1^2 \frac{\bar{Y}_k^2} { \bar{Y}_1^2} \right)
                   +  \langle\sigma_v^{kkk'k'}\rangle\left(Y_k^2- Y_{k'}^2 \frac{\bar{Y}_{k'}^2} { \bar{Y}_k^2} \right)\nonumber\\
                 &  +& \frac{1}{2} \langle\sigma_v^{kk23}\rangle\left(Y_k^2- Y_2Y_3 \frac{\bar{Y}_k^2} {\bar{Y}_2\bar{Y}_3}\right)
+    \langle\sigma_v^{2300}\rangle \left(Y_2Y_3-\bar{Y}_2 \bar{Y}_3 \right) 
     + \langle\sigma_v^{2311}\rangle \left(Y_2 Y_3- Y_1^2 \frac{\bar{Y}_2\bar{Y}_3} { \bar{Y}_1^2} \right)
\nonumber\\
&-& \frac{1}{2} \langle\sigma_v^{11k0}\rangle\left(Y_1^2-Y_k\frac{\bar{Y}_1^2}{\bar{Y}_k}\right) 
+ \langle\sigma_v^{k110}\rangle\left( Y_1 Y_k- Y_1 \bar{Y}_k \right)
\nonumber\\
 &+&\Gamma_{kk'}\left( Y_k -
Y_{k'}\frac{\bar{Y}_k}{\bar{Y}_{k'}} \right) - \Gamma_{k'k}\left( Y_{k'} -
Y_{k}\frac{\bar{Y}_{k'}}{\bar{Y}_{k}}\right) 
%+ \Gamma^{co}_{kk'}\left(Y_k -Y_{k'}\frac{\bar{Y}_k}{\bar{Y}_{k'}}\right) 
\label{eq:Y3}
\end{eqnarray}
where $k=2,3$, $k'=5-k$  and 
%$\Gamma_{kl}$ contains both the  decay and co-scattering terms~\cite{coscattering}
\begin{eqnarray} 
\nonumber
    \Gamma_{kk'}&=& \sum_{i\in k} g_i m_i^2  \frac{K_1(m_i/T)}{\sum_{j\in k} g_j m_j^2 K_2(m_j/T))}\Gamma^0_{i\to k',SM}
\end{eqnarray}

\noindent
where $\Gamma^0_{i\to k',SM}$ is the partial decay width for the decay for one DM into another DM and SM and $m_i$ are the DM masses. $K_i(x)$ are the Bessel functions of second order and degree $i$.

%P.S: we can still check the case where MDM1$<$MDM2. Here, A0 will freeze-out while S was still in T.E. (i.e Y1=Y1eq). At the T.F.O(A0), processes such as $22\rightarrow11$, $22\rightarrow00$ and $12\rightarrow10$ leads to a decreasing in omega2 value. As for omega1, as long as Y2 is large, processes such as $22\rightarrow11$ and $20\rightarrow11$ will increase omega1.

\subsection{Indirect detection}
\label{sec:ID}

The total number of events corresponding to DM pair annihilation is
 \begin{equation}
 N= \sum_{i,j} \rho^N_i \rho^N_j \langle v\sigma_{ij}\rangle (1-\frac{1}{2}\delta_{ij}) 
\end{equation}
where
\begin{eqnarray}
\rho^N_i &=& \rho\xi_i/m_i\\
\rho^N&=& \sum_i\rho^N_i
\end{eqnarray}
are respectively the number densities for DM$_i$ and the total DM number density.  $\xi_i$ is  the relative contribution of DM$_i$ to  the total DM density, $\xi_i=\Omega_i/\Omega_{tot}$, and  $\xi_1+\xi_2=1$ for two-component models. $\langle \sigma_v^{ij}\rangle$ is the short-hand notation for the cross-section for annihilation of DM$_i$DM$_j$ into all SM final states. We can define an effective annihilation cross-section 
\begin{equation}
    \langle \sigma_v\rangle =\frac{2N}{ {\rho^N}^2 }  
\end{equation}

To define the production rate of particles $a$ (here photons or antiprotons) from DM annihilation, we keep the same notation as for one-component DM,
\begin{equation}
 Q^a(E)= \frac{1}{2}  \langle \sigma_v \rangle {\rho^N}^2 \frac{dN^a(E)}{dE}
\end{equation}
where  the total yield reads
\begin{eqnarray}
 Q^a(E) &=&  \sum_{i>j}(1-\frac{1}{2}\delta_{ij})\rho^N_i\rho^N_j\langle \sigma_v^{ij}\rangle \frac{dN^a_{ij}(E)}{dE}\\ 
\end{eqnarray}
and $dN^a_{ij}/dE$ is the energy distribution of particle $a$ produced in DM$_i$DM$_j$ annihilation while $dN^a/dE$ is the effective spectrum for particle $a$.

The contribution of a given channel to the total cross-section, for example $Br(AAWW)$ for $\tilde{A}\tilde{A} \to WW$, relates the annihilation cross-section for that process to the total cross-section,

  \begin{equation}
   Br(AAWW) =  \frac{\langle \sigma_v^{AAWW} \rangle {\rho_2^N}^2}{\langle \sigma_v  \rangle {\rho^N}^2} 
     \end{equation}

\section{Constraints}
	\label{sec:constraints}
	
	\subsection{Theoretical Constraints}
\begin{figure}[h]
	\includegraphics[scale=0.5]{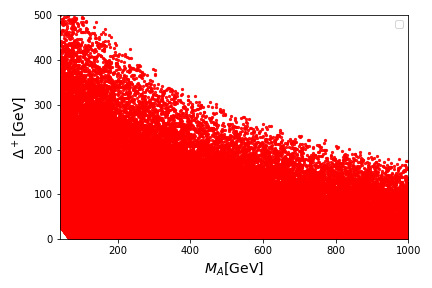}
	\includegraphics[scale=0.5]{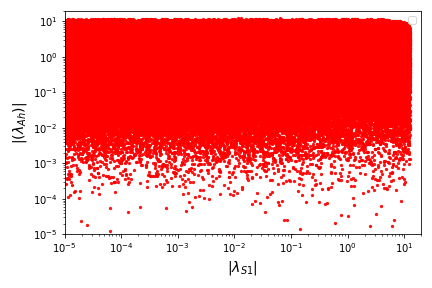}
	\caption{Impact of theoretical constraints, together with LEP and electroweak precision constraints on the mass difference $\Delta^+$  and on the couplings  of the two  DM to the Higgs, $\lambda_{Ah},\lambda_{S1}$}. 
	\label{fig:theo}
\end{figure}

\begin{itemize}
     \item{Perturbativity}
 We impose the condition that the vertex factor in the Feynman rules for quartic interactions  must be smaller than $4\pi$. This  ensures that the one-loop corrections are smaller  than the tree-level contributions ~\cite{Lerner:2009xg}.  This imposes an upper bound on the quartic  couplings, if a coupling enters several vertices we choose the strongest bound. This leads to the condition. 
  \begin{eqnarray}
  &&\lambda_2<\frac{2\pi}{3} ,\;   |\lambda_3| < 4\pi ,\;  |\lambda_3+\lambda_4| < 4\pi ,\;  |\lambda_4\pm\lambda_5| < 8\pi ,\;  |\lambda_3+\lambda_4\pm \lambda_5| < 4\pi ,\;  \nonumber\\
   && |\lambda_5| < 2\pi \;,\;\; |\lambda_{S1}|<4\pi  \;,\;\;    |\lambda_{S2}|< 4\pi \;,\;\; |\lambda_S|<\pi  \,,\;\;     |\lambda'_S|<\frac{\pi}{3} \;. 
  \end{eqnarray}
  
\item{Perturbative Unitarity}
For all scalar-scalar scattering processes at high-energy, the  partial wave unitarity condition should be satisfied for all scattering amplitudes.  We impose the condition that the eigenvalues of the scattering matrices  derived in ~\cite{Belanger:2014bga}  should all be smaller than $8 \pi$.

\item Stability conditions of the potential:\\

 To ensure that the quartic potential is bounded from  below, we write the matrix of quartic interactions in a basis of non-negative field variables (in the $r^2,s^2$ basis defined in ~\cite{Belanger:2014bga} ) and demand this matrix to be copositive.  Thus, we obtain the following general stability conditions 
\begin{equation}
\lambda_1>0 , \lambda_2>0, \lambda_S-\left|\lambda_S^\prime \right|>0 
\end{equation}
\begin{equation}
\lambda_3 + 2\sqrt{\lambda_1 \lambda_2}>0
\end{equation}
\begin{equation}
\lambda_3 + \lambda_4-\left|\lambda_5 \right| +  2\sqrt{\lambda_1 \lambda_2}>0
\end{equation}
\begin{equation}
\sqrt{\Lambda_{11} \Lambda_{22}} + \Lambda_{12}>0
\end{equation}
where
\begin{equation}
\Lambda_{11}=\lambda_1 \cos^4\gamma   + \lambda_2 \sin^4\left(\gamma \right) + \left(\lambda_3 + \left(\lambda_4+\lambda_{5} \cos\left( 2\Phi\right)\right)  \rho^2  \right) \cos^2\left(\gamma \right)\sin^2\left(\gamma \right)
\end{equation}
\begin{equation}
\Lambda_{22}=\lambda_S + \lambda_S^\prime \cos\left(4 \Phi_S \right) 
\end{equation}
\begin{eqnarray}
\Lambda_{12} &=&\frac{1}{2}\left[\lambda_{S1} \cos^2\gamma  + \lambda_{S2} \sin^2\gamma \right.  \nonumber\\
 &+& \left. \rho \sin \gamma \cos \gamma\left(\lambda_{S21} \cos\left(\Phi - 2 \Phi_S \right) + \lambda_{S12} \cos\left( \Phi + 2 \Phi_S\right)  \right)     \right] 
\end{eqnarray}
The last condition has to hold for all values of the parameters in the range $0\leq \gamma \leq \frac{\pi}{2}$, $0 \leq \left|\rho \right| \leq 1 $, $0 \leq \Phi \leq 2 \pi$ and $0 \leq \Phi_S \leq 2 \pi$. It is sufficient to compute the minimum of $\sqrt{\Lambda_{11} \Lambda_{22}} + \Lambda_{12}$ and check if it is positive. \\

\item{ Two stable dark matter candidates}

To  prevent the  decay channel $\tilde{A}^0 \rightarrow \tilde{S},\tilde{S}$ which is allowed for our choice of $Z_4$ charges, we take $M_A < 2 M_S$.

\item{Electroweak precision}

The electroweak precision parameters S,T are sensitive to physics beyond the SM,  in this model the one-loop contribution arise solely from the new scalar  doublet, thus are similar to the IDM,

\begin{equation}
 S=\frac{1}{72\pi}\frac{1}{\left( x_2^2 -x_1^2\right)^3 }\left[x_2^6f_a\left( x_2\right) - x_1^6f_a\left( x_1\right) +9x_2^2x_1^2\left(x_2^2f_b\left(x_2 \right)-x_1^2f_b\left( x_1\right)   \right)  \right]
\end{equation} 
where 
\begin{eqnarray}
f_a\left( x\right)&&=-5+12\log\left( x\right)\nonumber\\
f_b\left( x\right) &&=3-4\log\left( x\right) 
\end{eqnarray}
and $x_1=M_H/\MHC$, $x_2=M_A/\MHC$. When $x_1=x_2$, $S=\frac{1}{9\pi} \log(x_1)$,
\begin{equation}
T=\frac{1}{32\pi^2\alpha v^2}\left[f_c\left(\MHC^2, M_A^2 \right) +f_c\left(\MHC^2, M_H^2 \right) -f_c\left(M_H^2, M_A^2 \right)   \right] 
\end{equation}
where  
\begin{eqnarray}
f_c\left( x,y\right) &&=\frac{x+y}{2} - \frac{xy}{x-y}\log\left(\frac{x}{y} \right), \hspace{12pt} x\neq y\nonumber\\
f_c\left( x,y\right)&& = 0, \hspace{12pt} x=y
\end{eqnarray}

Assuming a SM Higgs boson mass  $m_{h}=125$ GeV,  the  values of the S and T parameters in the limit U=0, have been determined to be 
\begin{equation}
S=0.06\pm0.09, \hspace{12pt} T=0.1\pm 0.07
\end{equation}
with a correlation coefficient +0.91\cite{Baak:2014ora}.  The EWPT data prefers a modest value for  the  mass splitting between the charged and neutral components of the doublet, roughly 
$\Delta^+ < 500 \rm{GeV}$ as can be seen in Fig.~\ref{fig:theo}. When scanning over the parameter space we have computed S and T for each point and have required that they  fall within the ellipse corresponding to the 95\% C.L. limit. 
\end{itemize}

\subsection{Collider Constraints :  LEP} 
  
In the low mass range, LEP experiments constrains the parameter space of the model, these limits are similar to the ones obtained in the IDM,
\begin{itemize} 
 \item Widths of $W$ and $Z$ bosons\\ 
The precise measurements of the $W$ and $Z$ widths  imposes that the decay of the gauge bosons,  $W^+\rightarrow \tilde{H}_0\tilde{H}^+, \tilde{A}_0\tilde{H}^+$ and $Z\rightarrow \tilde{A}_0\tilde{H}_0, \tilde{H}^+\tilde{H}^-$ 
should be kinematically forbidden, thus 
\begin{equation}
M_A + \MHC > M_W ,\hspace{12pt} M_H + \MHC > M_W
\end{equation}
\begin{equation}
M_H+ M_A >M_Z , \hspace{12pt} 2\MHC > M_Z
\end{equation}

\item Dijets and di-leptons\\
The process $e^+e^- \rightarrow \tilde{H}^0\tilde{A}^0$  could lead to a visible di-jet or di-lepton signal from $\tilde{H}\to \tilde{A} \bar{f}f$, a reinterpretation of searches for neutralino pair production in Ref.~\cite{Lundstrom:2008ai} rules out   the region defined by the intersection of the conditions below
    \begin{equation}
    M_H<80 {\rm GeV} \;,\;\;M_A <100 {\rm GeV}\;,\;\; {\rm if} \;\;\; M_A - M_H>8 {\rm GeV}
    \end{equation} 
\item Charged Higgs\\
The process  $e^+e^- \rightarrow \tilde{H}^+\tilde{H}^-$ at LEP2 sets a limit $\MHC>70 GeV$
%\begin{equation}
%\MHC>70 GeV
%\end{equation}
 as a result of the re-interpretation of limits on charginos \cite{Pierce:2007ut}. The lower limit on the charged Higgs is more stringent  in the case of a compressed spectra. If the  charged Higgs is long-lived with a lifetime $\tau>10^{-6}$ sec, searches for heavy stable charged particles further restricts the charged Higgs to be $\MHC>100$GeV~\cite{OPAL:2003zpa}. Moreover it also constrain lighter charged Higgses if their lifetime is such that 95\% of them have decayed before 3m.  For $\MHC=70$~GeV this corresponds to $\tau>3.1\times 10^{-9}$sec.
\end{itemize}

By applying theoretical constraints (stability and unitarity) as well as  Electroweak Precision Test limits and LEP constraints for the general scan, leads to an upper limit on the mass difference within the scalar doublet, Fig.~\ref{fig:theo} (left), while the couplings of the two DM to the Higgs cover the whole range used in the scan,  Fig.~\ref{fig:theo} (right).

\subsection{Collider Constraints : LHC} 
\label{sec:LHC}

Precise measurements of the properties of the Higgs at the LHC further constrain the scalar sector, only two channels can be affected in the Z4IDSM.

    \begin{itemize}
    	\item Invisible decay of the Higgs \\
	 For light DM, the SM  Higgs can decay into two DM scalar particles.  The direct $95\%$ C.L. limit   on the invisible width was set by CMS at 19\%~\cite{CMS:2018yfx} using the vector boson fusion process while a more stringent limit  was obtained by  ATLAS by combining searches for an invisible Higgs produced in vector fusion or in association with top quarks at 13 TeV with searches at 7 and 8 TeV \cite{ATLAS_invisible}, we will use the latter limit, 
    	\begin{equation}
    	Br\left( h \rightarrow invisible\right)< 11\% .
    	\end{equation}  
    	\item The di-photon decay rate\\
	The  charged Higgs contributes to the one-loop induced process  $h \rightarrow \gamma \gamma$ decay. The corresponding partial decay width is computed within \micro~\cite{Belanger:2013oya} and this is compared with the experimental upper bound using HiggsSignals~\cite{Bechtle:2020uwn}.

    %	\begin{equation}
   % 	\frac{Br^{BSM}\left(H_1 \rightarrow \gamma \gamma \right) }{Br^{SM}\left(H_1 \rightarrow \gamma \gamma \right) }= 1.14^{+0.38}_{-0.36} 
   % 	\end{equation}
   % 	at $95\%$ CL.
    \end{itemize}
Finally new physics searches at the LHC allow to probe the model, those are monojet searches, disappearing track searches, searches for heavy stable charged particles (HSCP) and dileptons.

\begin{itemize}
\item ATLAS monojet search\\
A recast of the monojet search was also performed in ~\cite{Belyaev:2018ext} and ~\cite{Belyaev:2016lok} in the IDM. The exclusion in the $\lambda_{Ah} -M_{DM}$ plane for different luminosities shows a sensitivity around $\lambda_{Ah}=0.03$ for $M_{DM}<60$GeV which degrades to ${\cal O}(1)$ as soon as the SM invisible Higgs threshold is crossed. We have used the exclusion for 20 - 100 fb$^{-1}$ as well as the projection for 3ab$^{-1}$. They have little impact on the parameter space of the Z4IDSM model. We have also estimated the contribution of the singlet scalar which for $M_S=M_A$ and $\lambda_{S1}=\lambda_{Ah}$ is twice that for the pseudoscalar doublet.  As we will see in the following, the higher mass and the different range of couplings allowed entail that the singlet does not contribute to present or future monojet searches at the LHC. 
An analysis of the monoZ and monoH processes in ~\cite{Belyaev:2016lok} also finds that only the region where $M_H<60$GeV could be probed even with the high luminosity (3ab$^{-1}$). Thus we ignore these two processes in the following.

\item $jj +MET$ from Vector Boson fusion\\
In the inert doublet model, a recast  of a search for an invisible Higgs produced through vector boson fusion (VBF) from the CMS collaboration in Run2~\cite{CMS:2018yfx} was performed in Ref~\cite{Dercks:2018wch}. The VBF processes  was found to be more constraining than the above mentionned monojet search from a recast of the ATLAS collaboration  result ~\cite{ATLAS:2017bfj} based on CHECKMATE~\cite{Drees:2013wra,Dercks:2016npn}. The VBF process rules out only the region   $\lambda_{Hh}>1(3)$ for $M_H\approx 65(70)$ GeV and only  $\lambda_{Hh}>10$ for $M_H\approx 85-100$ GeV. We expect similar limits to be obtained on $\lambda_{Ah}$ in the case where the pseudoscalar is the DM. In the following we will see that such large values of $\lambda_{Ah}$ are not allowed by other constraints in the Z4IDSM model, thus this process does not restrict further the parameter space of the model. We also checked that additional contributions involving the singlet scalar were always much suppressed relative to the doublet contribution, hence would not give additional constraint.

\item Long-lived particles \\
When the mass splitting between the charged Higgs and DM is small (at most few hundred MeV's), the charged Higgs is long-lived. There are two collider searches aimed at long-lived particles  that can be relevant in this model : the search for heavy charged particles (HSCP) and the disappearing track search which correspond to the charged Higgs decaying into soft particles and DM with $c\tau\approx m$. To compute reliably the lifetime of the charged Higgs we modify the model to include the effective vertices $\tilde{H}^+\tilde{A}^0 \pi^-$ and $\tilde{H}^+\tilde{H}^0 \pi^-$ as described in section 2.  Including pions in the decay mode decrease significantly the lifetime of the charged Higgs. This effective vertex is taken into account only if the mass splitting lies in the range $140 MeV\leq \Delta^+ \leq 500 MeV$. For larger mass splittings a complete calculation would also include the decay into two pions, we have however  found that even without this mode the lifetime of the charged Higgs was short enough to be outside the range of applicability of the disappearing track, see Fig.~\ref{fig:ctau_deltaM}. For our analysis a precise knowledge of the charged Higgs lifetime is not required. 
We implemented the limits that were obtained in ~\cite{Belyaev:2020wok} for disappearing tracks. We also checked a posteriori that these limits were consistent with the ones included in the latest version of SModelS2.0.0~\cite{Ambrogi:2018ujg,Khosa:2020zar,Alguero:2020grj}. 
For very small mass splitting  we use the limits from HSCP searches based on ~\cite{CMS:2015lsu,ATLAS:2019gqq} that  were  implemented via SModelS2.0.0~\cite{Heisig:2018kfq,Ambrogi:2018ujg}.

\item Dilepton searches\\
Searches for opposite-sign dileptons and $E_T^{miss}$ at the LHC can constrain the IDM model. In this model three processes contribute to dileptons, $pp\rightarrow \tilde{A}\tilde{H},\tilde{H}^+\tilde{H}^-,Z\tilde{H}\tilde{H}$, where dileptons come from the decay of the Z or from the decay $\tilde{H}^+\rightarrow \tilde{A}W^+$.The signature is similar to the one of neutralino/chargino in supersymmetry. A study of Run1 ~\cite{Belanger:2015kga} based on a recast of the SUSY search ~\cite{ATLAS:2014zve} showed that only the region where the lightest element of the doublet is lighter than 60 GeV was constrained, the limit depending on the mass of other members of the doublet. In the Z4IDSM model, we expect similar results. Note that because these searches are targeted at supersymmetry the cuts applied may not be optimal for the IDM model,  a study showed~\cite{Chakraborti:2018aae} that in the IDM the $E_T^{miss}$ peaked at much lower energy such that a softer cut on $E_T^{miss}$ has the potential to increase the sensitivity in the IDM model, thus to the Z4IDSM model. Moreover analyses of the IDM at the  high luminosity LHC have shown that multilepton channels could probe the model when the doublet DM is below the electroweak scale ~\cite{Datta:2016nfz}.

\begin{figure}[h]
	\includegraphics[scale=0.5]{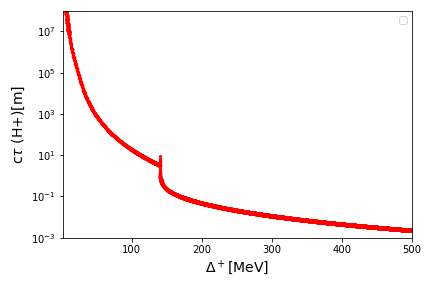}
	\centering
	\caption{$c\tau$  as a function of the mass splitting $\Delta^+=M_{H+}-M_A$.}   
	\label{fig:ctau_deltaM} 
\end{figure}

\end{itemize}

\subsection{Dark Matter constraints}

\noindent
{\bf Relic density } 

In this study we will require that both DM components explain DM, thus we require that the total relic density falls within the observed range determined
very precisely by the PLANCK collaboration to be $\Omega h^2= 0.1184 \pm 0.0012$~\cite{Planck:2015fie}. We also take into account a 10\% theoretical uncertainty, this rough estimate is motivated by previous studies performed  in the inert doublet model,  a model which shares many of the characteristics of the $Z_4$IDSM, which showed that  the full one-loop corrections to the relic density were  typically of that order  ~\cite{Banerjee:2019luv,Banerjee:2021hal}.  We take the $2\sigma$ range  that is we consider the following range for the total  $\Omega_{DM} h^2 =\Omega_1 h^2+\Omega_2 h^2$.
	\begin{equation}
	0.094<\Omega_{DM} h^2 < 0.142
	\end{equation}
	
\noindent	
{\bf Direct Detection }

 Scalar DM interacts with nuclei only through spin-independent (SI) interactions. For DM masses near  the electroweak scale,  as considered  here, the best limits on the SI cross-section for DM scattering on nucleons have been obtained  by XENON1T~\cite{Aprile:2018dbl}. In the Z4IDSM, we compute the recoil energy distribution including  the two DM signals scaled by their fractional contribution to the total DM density,  and use the recasted limits at 90\%C.L.  implemented in \micro~\cite{Belanger:2020gnr}. For this we use the default values for the astrophysical parameters, namely a local DM density $\rho_{DM}= 0.3 {\rm GeV/cm}^3$, and a Maxwellian velocity distribution with the following parameters:  the rotation velocuty of the Galaxy, $v_{Rot}=220{\rm km/s}$, the escape velocity of the Galaxy $v_{esc}=544 {\rm km/s}$, the velocity of the Earth in the galactic plane
 $v_{Earth}=232 {\rm km/s}$.\\
	
\noindent	
{\bf Indirect Detection } 	

Fermi-LAT observations of photons from Dwarf Spheroidal Galaxies (dSph), provide one of the most robust constraint on DM especially at masses below the electroweak scale when DM annihilates into fermion pairs ($bb$ or $\tau\tau$). In our model, as will be discussed later, the DM annihilation cross-sections  are in general largest for the gauge bosons final states.  Thus we compute $\langle \sigma_v \rangle$ for  $\tilde{A}\tilde{A}$ into  WW and ZZ final states  and require that  
$\langle \sigma_v \rangle_VV\equiv \langle \sigma_v^{AAVV} \rangle=\langle \sigma_v^{AAWW} \rangle +\langle \sigma_v^{AAZZ} \rangle$
 lies below the 95\%C.L. given by FermiLAT in ~\cite{Fermi-LAT:2015att}.  Note that the photon spectrum from ZZ final states is slightly below that of WW, we ignore this difference. This approach is conservative as other channels can contribute as well as will be discussed in Section~\ref{sec:sub:ID}.

Searches for anti-protons performed by AMS-02~\cite{AMS:2016oqu} provide the most severe constraints on  WIMP DM.  However these constraints strongly depend on the CR propagation parameters as well as on the DM profile. The uncertainty on the cosmic ray (CR) propagation parameters have been greatly reduced by AMS-02 measurements of the B/C ratio~\cite{Genolini:2021doh,AMS:2016brs}. Here again we adopt a conservative approach and include only the gauge bosons final states. For this we use the limits derived in ~\cite{Reinert:2017aga} for WW final states, and apply it to both the WW and ZZ final states,  the antiproton spectrum from WW and ZZ being close to each other. In ~\cite{Reinert:2017aga} the size of the diffusion halo is set to the minimum value $L=4.1kpc$ and  the remaining propagation propagators and the exclusion cross-section in the WW channel for each DM mass are obtained from a global fit to B/C and to the antiproton spectrum. The fit  is done for three DM profiles, we will use the most conservative limit obtained assuming a generalised NFW profile with $\rho=0.3 {\rm GeV/cm}^3$ and also display the limit for a standard NFW profile with $\rho=0.38 {\rm GeV/cm}^3$.

Limits on DM annihilation  will also be obtained by CTA  which  measures the photon spectrum albeit at higher energies~\cite{Acharyya:2020sbj}. We perform a dedicated analysis to determine the parameter space of the model that is  within reach of CTA. For this we use the combined photon spectra from all annihilation channels as described in section~\ref{sec:ID}.

\section{Results}
\label{sec:results}

To determine the parameter space of the model compatible with  all current theoretical and experimental constraints, we have performed random scans over the masses and couplings in Eq.~\ref{eq:free}. First using a wide range for all parameters (the general scan) then in the second case restricting to the region where the doublet is nearly degenerate in mass, although this case requires to impose the conditions that some parameters are very small, the different DM and collider phenomenology warrant a separate investigation.

\subsection{General scan}
The free  parameters of this model are varied randomly  using a logarithmic  scan for the couplings and a linear scan for the masses and/or mass splittings. Since we have already shown that the theoretical constraints, including EWPT, require that the mass differences among the components of the doublet cannot be too large,  Fig.~\ref{fig:theo}, we used as free variables the mass of the doublet  DM $M_A$ as well as
the mass differences, $\Delta^0$ and $\Delta^+$. The range of the free parameters are 
  given in Table~\ref{tab:range}.
% for the coupling we use log.
 \begin{table}[!htb]
\begin{center}
\begin{tabular}{|cc|cc|cl|}
\hline
 $M_A$ & $40 - 1000$ GeV & $|\lambda_{S1}|$&$10^{-5} - 4\pi$ &  $\lambda_2$&$10^{-5} - 2\pi/3$\\
    $\Delta^0$  &$0 - 500 $ GeV& $|\lambda_{S2}|$&$10^{-5} - 4\pi$  &  $|\lambda_3|$&$10^{-5} - 4\pi$\\
    $\Delta^+$  &$0 - 500 $ GeV&  $|\lambda_{S12}|$&$10^{-5} - 4\pi$    &  $\lambda_S$&$10^{-5} - \pi$\\
    $M_S$   &$40 - 1500$ GeV& $|\lambda_{S21}|$&$10^{-5} - 4\pi$&$|\lambda'_S|$&$10^{-5} - \pi/3$\\  
     \hline
     \end{tabular}
     \caption{Range of the free parameters of the Z4IDSM model used in the scan.}
     \label{tab:range}
     \end{center}
\end{table}

\subsubsection{Relic density and direct detection}

\begin{figure}[h]
	\includegraphics[scale=0.5]{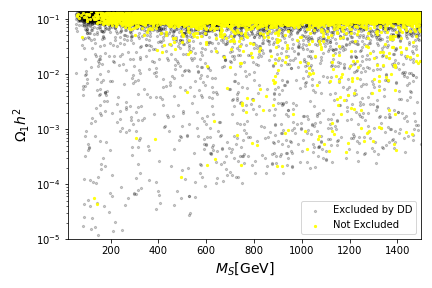}
	\includegraphics[scale=0.5]{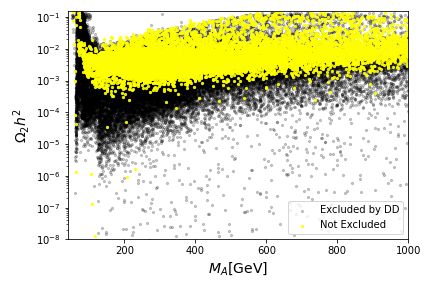}
	\centering
	\caption{$\Omega_1h^2$ (left) and $\Omega_2h^2$ (right) as a function of the corresponding  DM mass, for all points satisfying theoretical, collider and the total relic density constraints (grey) as well as DD constraint from XENON-1T(yellow). }   
	\label{fig:omega12} 
\end{figure}

The relic density constraint from PLANCK and  the direct detection limits from XENON1T provide  important constraints on the model. 
We find that generally the singlet forms   the dominant DM component while the doublet can be dominant only in a narrow region around $m_h/2$ and for masses $M_A>500$GeV, see Fig.~\ref{fig:omega12}. These two regions correspond to the ones where in the IDM  the DM relic density is compatible with the PLANCK measurement. Moreover since a large doublet contribution to the  relic density requires a compressed spectra, this region is only sparsely populated. We will discuss the case of the compressed spectrum with a dedicated scan in the next section. We find that DD severely restricts the allowed parameter space and that in particular
the region  where $m_A\approx m_h/2$ is strongly constrained by DD. For most of the allowed points  the singlet and doublet  components each contribute at least to 1\% of the total relic density. In a few cases the doublet component is at the per-mil level or below, notably we found a few cases for $M_A<250 GeV$ where $\Omega_2 h^2 \approx 10^{-6}$. 
These correspond to specific configurations where important DM conversion or semi-annihilation  reduces both  $\Omega_1 h^2$ and $\Omega_2 h^2$   when $M_A >M_S$, or 
scenarios where  the doublet or the singlet lies near the Higgs resonance and require a very weak coupling to the Higgs that allows to  escape DD constraints.  After imposing the relic density and DD constraints from XENON-1T in addition to theoretical and collider constraints, the range of allowed masses for the two DM covers almost the full range probed, the lower limit on the singlet and doublet being respectively $M_S > 58.3$GeV and $M_A>56.3$GeV.  In particular  the singlet can form most of the DM and satisfy DD constraints  for the full mass range, in stark contrast with the singlet DM model.We have also checked that, as expected, the region where $M_A>2M_S$ has the same properties as the singlet model, with the only allowed region for $M_S\approx m_h/2$ or above the TeV scale. 
	
\begin{figure}[hbt]
	\includegraphics[scale=0.45]{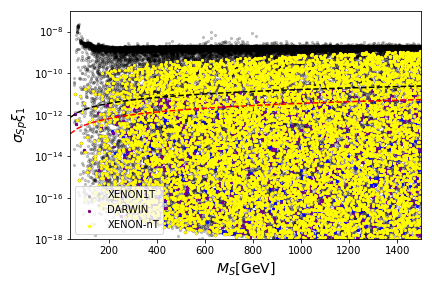}
	\includegraphics[scale=0.45]{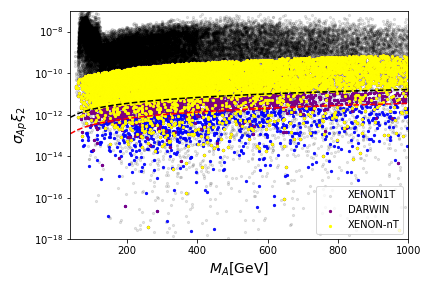}
	\centering
	\caption{ Spin-independent DM proton scattering cross section   times the fraction of DM component,  as a function of the corresponding DM masses, $\sigma_{Sp} \xi_1 - M_S$ (left) and $\sigma_{Ap} \xi_2 -  M_A$ (right). The black (red) line shows the projected reach of XENON-nT (DARWIN), points in yellow (red) are within the reach of XENON-nT (DARWIN), and blue points are beyond the reach of DARWIN. }   
	\label{fig:sigma} 
\end{figure}

To implement the DD constraint from XENON-1T, we recall that we have combined the signal from both DM.  We first  remark that even if $\tilde{A}$ is subdominant, it is still effectively probed by XENON1T, see Fig.~\ref{fig:sigma} where we show the predictions for the spin independent cross-section  for each DM candidate over the allowed parameter space. Each contribution is rescaled by  the appropriate fraction of the DM candidate,
$\xi_i$.   On the same figure we also make a naive projection of the reach of XENON-nT and DARWIN, where for simplicity we consider each DM signal separately. The vast majority of points are within the reach of XENON-nT  for the doublet DM, however in some cases the signal is suppressed by two or three orders of magnitude. Points in red that lie below the line showing the XENON-nT projection in Fig.~\ref{fig:sigma} left (right)  are within reach of XENON-nT for the doublet (singlet)  component. Clearly the singlet component often easily escapes DD even if it forms the dominant DM.  The presence of the doublet and of the conversion/semi-annihilation channels allow to reduce the value of $\lambda_{S1}$ required by the relic density constraint, thus escaping current and future DD searches.

\begin{figure}[htb]
	\includegraphics[scale=0.5]{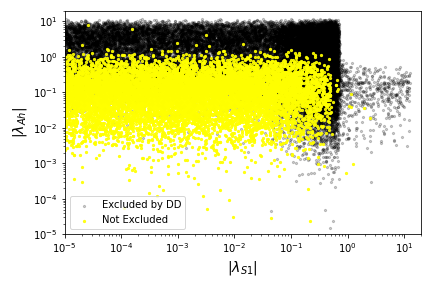}
	\includegraphics[scale=0.5]{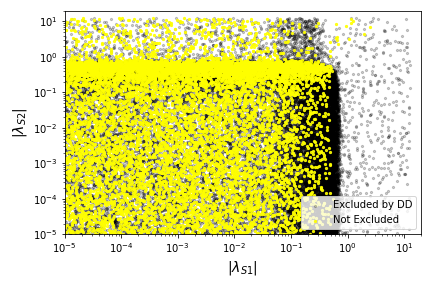}
	\centering
	\caption{Allowed points in the $\lambda_{Ah}$ - $\lambda_{S1}$ plane (left)   and  $|\lambda_{S2}| - |\lambda_{S1}|$ plane (right). Points in black are excluded by XENON-1T. }   
	\label{fig:semi_conv} 
\end{figure}

\begin{figure}[htb]
	\includegraphics[scale=0.5]{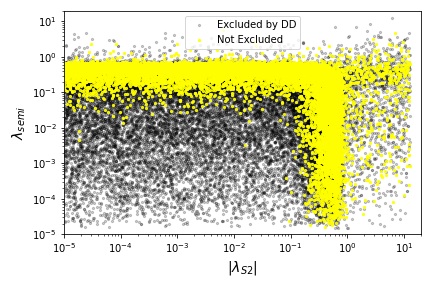}
	\includegraphics[scale=0.5]{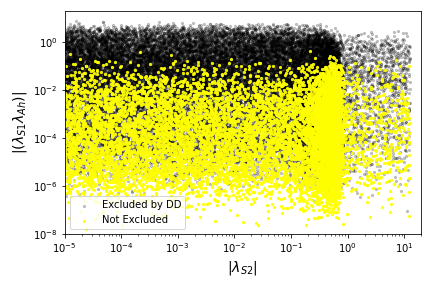}
	\centering
	\caption{Semi-annihilation couplings: allowed points in the $\lambda_{semi}$ - $\lambda_{S2}$ (left)  and $\lambda_{Ah}  \lambda_{S1}$ - $\lambda_{S2}$ plane (right), same color code as Fig.~\ref{fig:semi_conv}. }   
	\label{fig:semi} 
\end{figure}

The impact of the relic density and XENON-1T constraints on the free couplings of the model are illustrated in Fig.~\ref{fig:semi_conv} and ~\ref{fig:semi}. The couplings that are responsible for DM formation in the singlet and IDM models,   $\lambda_{S1}$ and  $\lambda_{Ah}$ respectively, cover a much wider range than in each of the sub-models, see Fig.~\ref{fig:semi_conv} (left panel). In particular, contrary to the singlet model, $\lambda_{S1}$ can be as small as $10^{-5}$, the lower limit in our scan. In that case, as mentioned above,  interactions between the two DM  govern the annihilation of the singlet component.  \footnote{Note that since we performed a random scan with a linear dependence on the mass differences between the singlet and the doublet components, we have a sparse sampling of very compressed spectrum, hence we have a sparse sampling of the small values of $\lambda_{Ah}$. This case will be covered in Section ~\ref{sec:ND}.}  This means that small values of $\lambda_{S1}$   are  associated with either a large $\lambda_{S2}$ coupling, which determines DM conversion or a large $\lambda_{semi} = \sqrt{\lambda_{S21}^2+\lambda_{S12}^2}$ responsible for semi-annihilation.

Direct detection also strongly impacts the allowed parameter space.
 As expected,  the largest values of the  couplings  $\lambda_{Ah}$ and $\lambda_{S1}$ responsible for the SI cross-section of the doublet and singlet components respectively  will be excluded, as shown in Fig. ~\ref{fig:semi_conv} (left panel). As concerns other couplings, 
 the region where both $\lambda_{semi}$ and  $|\lambda_{S2}|$ are small, which is allowed by theoretical constraint and relic density, is ruled out by  DD constraint.  Indeed  if semi-annihilation and DM conversion are negligible, the large value of $\lambda_{S1}$ required by the relic density constraint is excluded by DD, unless $M_S$ is above the TeV scale. We therefore conclude that some amount of semi-annihilation or DM conversion is necessary to satisfy both relic density and direct detection. 
Note that the product of couplings $\lambda_{Ah} \times \lambda_{S1}$ can also lead to DM conversion through processes of the type $\tilde{S}\tilde{S}^\dagger\rightarrow \tilde{H}\tilde{H}$, however these couplings are constrained by DD to be  $\lambda_{Ah} \times \lambda_{S1}<5\times10^{-2}$, see Fig.~\ref{fig:semi} (right). Thus in general DM conversion is dominated by $\lambda_{S2}$. 
After applying all current constraints, we also note that the bulk of the points are in the region where $0.1<\lambda_{S2}<1$ with extension all the way to the upper limit. Similarly,
$0.1< \lambda_{semi}<1$, these conditions are required to suppress the abundance of the singlet without having a too large $\lambda_{S1}$. In Fig.~\ref{fig:semi_conv}right,  we clearly see that large values of  $\lambda_{S1}$ are disfavoured although $\lambda_{S1}$ can reach 1 when  it is associated with a rather heavy DM  - thus evading DD constraints.

	\begin{figure}[htb]
	\includegraphics[scale=0.5]{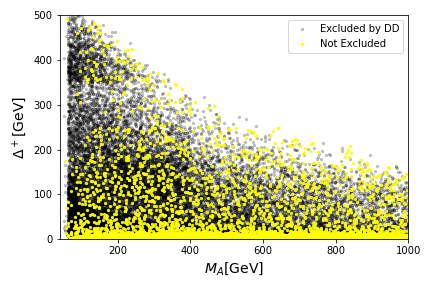}
	\includegraphics[scale=0.5]{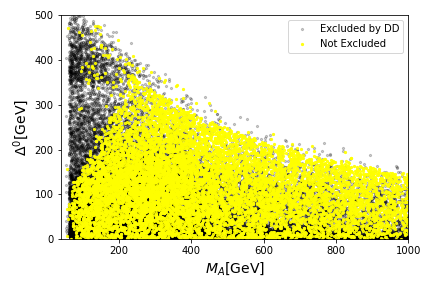}
	\centering
	\caption{Allowed points in the $\Delta^+$ - $M_A$ (left)  and $\Delta^0$ - $M_A$ plane (right), same color code as Fig.~\ref{fig:semi_conv} }   
	\label{fig:deltam} 
\end{figure}

\subsubsection{Indirect detection : photons and antiprotons}
\label{sec:sub:ID}

First we summarise the contributions from the different final states for each pair of DM, for this we compute
  $\langle \sigma_v^{ijkl} \rangle$   for  which we use  a shorthand  notation  $\langle \sigma v \rangle_{kl}$  representing  the contribution of a given initial state of DM particles in the channel $kl$ to the total annihilation cross-section. This quantity implicitly takes into account the fraction of the DM abundance for each DM candidate.
In the corresponding figures we display only final states for which the cross section can reach at least  $10^{-27}{\rm cm}^3/{\rm s}$ since this is already below the sensitivity of Fermi-LAT and even of  the future CTA. 
 
\begin{figure}[htb]
	\includegraphics[scale=0.4]{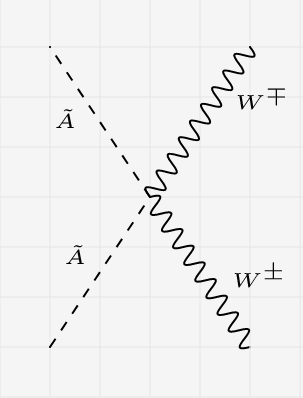}
	\includegraphics[scale=0.4]{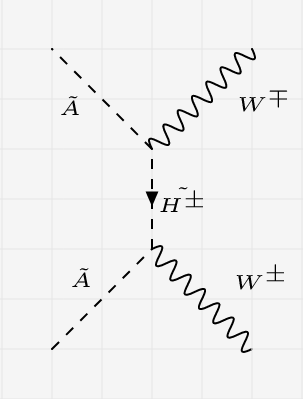}
	\centering
	\caption{ Feynman diagrams that contribute to $\tilde{A}\tilde{A}\to W^+W^-$ }   
	\label{fig:feynman} 
\end{figure}

\begin{figure}[h]
	\includegraphics[scale=0.45]{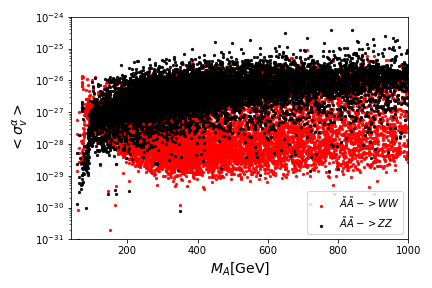}
	\includegraphics[scale=0.45]{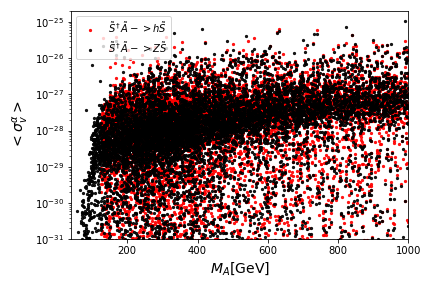}
	\includegraphics[scale=0.45]{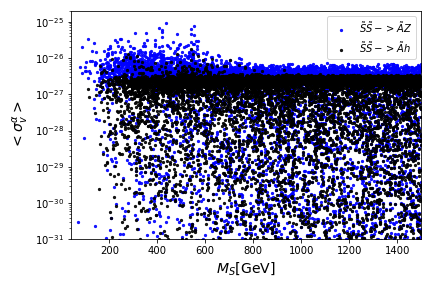}
	\includegraphics[scale=0.45]{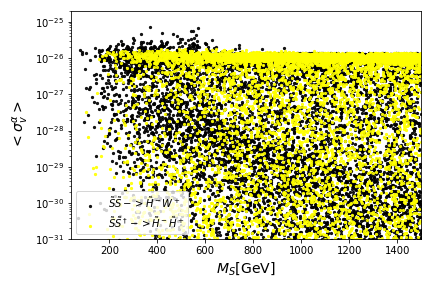}
	\centering
	\caption{ Cross-section, $\langle \sigma_v^\alpha\rangle$  for the channels    that lead to the largest cross-sections. The channels $\alpha$ are specified in each panel. }   
	\label{fig:sigmav:channels} 
\end{figure}

Despite the fact that $\tilde A$ is the subdominant DM component, the indirect detection signal from pair annihilation into W(Z) pairs can be dominant, it can reach $10^{-25} {\rm cm}^3/{\rm s}$, see Fig.~\ref{fig:sigmav:channels}.  Moreover,  $\langle \sigma_v \rangle_{WW/ZZ}$  can exceed $10^{-26} {\rm cm}^3/{\rm s}$ in almost the full range of $M_A$ considered, the region where the cross section is suppressed correspond to the one below roughly 100GeV. Note also that in many cases the $ZZ$ channel is much larger than the $WW$ channel. This peculiar behaviour is related to the fact that when the doublet is compressed, there is a cancellation between the quartic diagram and the t-channel diagram,  Fig.~\ref{fig:feynman} leading to a much smaller cross-section. Indeed,   the mass splitting  $\Delta^+$ is generally small, see Fig.~\ref{fig:deltam},  hence leading to a suppressed cross-section in the WW channel   while the mass splitting  $\Delta^0$ between neutral components can be much larger, hence the ZZ  channel  is less affected by the cancellation between the two diagrams.
 The main subdominant contribution is from the hh channel, the largest value being  $\langle \sigma_v \rangle_{hh} = 4\times 10^{-27}  {\rm cm}^3/{\rm s}$ while in general the quark channels (tt or  bb)  are  suppressed by  many orders of magnitude.  For example, the latter reaches at most  $4 \times 10^{-28}  {\rm cm}^3/{\rm s}$.% while the $t\bar{t}$ channel reaches at most $2.8\times 10^{-27}cm^3/sec$. 

The annihilation cross-section  for the dominant DM component, $\tilde{S}$, into pairs of SM particles is much smaller than for the doublet,  it can reach  only $7\times 10^{-27} {\rm cm}^3/{\rm s}$.  The channels with larger cross-sections are WW,ZZ and to a lesser extent hh, while  $t\bar{t}$ and $b\bar{b}$ are negligible, lying below $10^{-28}  {\rm cm}^3/{\rm s}$. 
Interactions between the two DM components often provide the dominant semi-annihilation channels for the singlet.  For semi-annihilation processes such as  $\tilde{S}\tilde{S}\rightarrow \tilde{H}^-W^+$  or $\tilde{H}^+W^-$ the cross-section is typically concentrated near  $10^{-26} {\rm cm}^3/{\rm s}$ for any value of $M_S$  and can in a few cases reach $7\times10^{-26} {\rm cm}^3/{\rm s}$. Other channels where a pair of singlet DM annihilates into a doublet component and a standard model, such as
$\tilde{S}\tilde{S}\rightarrow  \tilde{A}h, \tilde{A}Z$   can exceed $10^{-26} cm^3/sec$ especially for $M_S<500$GeV.  Note that
 $\tilde{S}\tilde{S}\rightarrow  \tilde{H}h, \tilde{H}Z$ lead to similar cross-sections after having taken into account the heavier mass of $\tilde{H}$. Semi-annihilation processes such as   $\tilde{S}^\dagger\tilde{A}\rightarrow h\tilde{S},Z\tilde{S}$  lie generally below $10^{-26}cm^3/sec$ except for a few points.

The DM conversion processes can also be important, notably $\tilde{S}\tilde{S}\rightarrow \tilde{H}^+\tilde{H}^-$ for which $\langle \sigma_v \rangle_{H^+H^-}$  is generally around $10^{-26}cm^3/sec$ for any masses  while   $\tilde{S}\tilde{S}\rightarrow \tilde{H}^0\tilde{H}^0$  is slightly suppressed.  Finally other DM conversion processes such as   $\tilde{A}\tilde{A}\rightarrow \tilde{S}\tilde{S}$  or the reverse process can be large, reaching above $10^{-25} cm^3/sec$,   however these modes are invisible and  do not contribute to the signal. 
 
\begin{figure}[hbt]
	\includegraphics[scale=0.55]{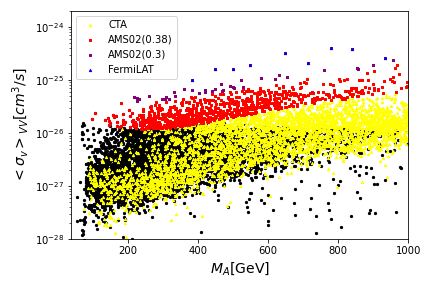}
	\centering
	\caption{ Points  excluded by FermiLAT (blue)  and AMS02 assuming NFW and $\rho=0.3$ (purple) or $\rho=0.38$ (red) in the $\langle \sigma_v\rangle_{VV}$-$M_A$ plane. The reach of CTA is also shown (yellow), points in black are beyond the reach of these detectors.  }   
	\label{fig:fermi_ams} 
\end{figure}

After having established that the  cross-sections that are in the realm of FermiLAT  are to a large extent only possible in the gauge bosons channel,   to impose current constraints from Fermi-LAT we  include only the VV channels where we sum over the W and Z final states.  We find that  WW/ZZ channels exclude  the region with the largest cross-sections  for large $M_A$, the scenarios that are currently exclude by FermiLAT are displayed in Fig.~\ref{fig:fermi_ams}. 

 To impose the constraint from AMS02, we follow the analysis in ~\cite{Reinert:2017aga} which gives the excluded cross-section from the WW channel assuming different profiles. We assume the NFW and a cored generalised NFW profile. The latter leading to the most conservative limit.   The scenarios that are currently exclude by AMS02 under these assumptions are also displayed in Fig.~\ref{fig:fermi_ams}.  Clearly the uncertainty on the limits coming from the assumption on the profile is large, this further justifies our approximation of similar spectrum from ZZ and WW channels and of neglecting the semi-annihilation channels. Note that values of $M_A$ up to 1 TeV (the upper limit in our scan) can be probed by AMS02 for the largest cross-sections.

Next we determine whether the  increased sensitivity of future detectors might allow to probe scenarios where $\langle \sigma_v \rangle$ is dominated by semi-annihilation and or DM conversion processes.  First we illustrate the impact of non-SM channel on the photon spectrum using the two benchmarks in Table~\ref{tab:bench}. For these the dominant contributions to $\langle \sigma_v \rangle$ come from non-SM final states. 
For the first benchmark, the semi-annihilation couplings are of ${\cal O} (1)$ hence semi-annihilation form the dominant channels, with $\tilde{S}^\dagger\tilde{A}\to Z\tilde{S}$ (44\%), $\tilde{S}\tilde{S}\to W^-\tilde{H}^+$(20\%) and $\tilde{S}\tilde{S}\to W^+\tilde{H}^-$(12\%),  $\tilde{S}\tilde{S}\to Z\tilde{A}$(18\%) and $\tilde{S}^\dagger \tilde{A}\to h\tilde{S}$ (6\%). The total cross-section $\langle \sigma_v \rangle =7 \times 10^{-26}{\rm cm}^3/{\rm s}$ and for all energies except for $E_\gamma>200$ GeV, the full spectrum is dominated by $\tilde{S}^\dagger\tilde{A}\to Z\tilde{S}$.
Moreover we find that the photon spectrum for channels with $\tilde{H}^\pm W^\mp$ final state is very soft. For this benchmark, the spectrum is such that $\tilde{H}^\pm W^\mp$ cannot be produced on-shell, moreover the mass splitting between the charged and neutral Higgs do not allow for the charged Higgs to decay into a real W, rather the decay proceed via the 3-body final state, thus the photon spectrum is soft. 
For the second benchmark, the spectrum is slightly heavier and the mass difference $\Delta^+$ is only a few GeV's, as just argued the photon spectrum associated with the dominant channel   $\tilde{S}\tilde{S}\to W^-\tilde{H}^+$ (46\%)  is very soft. With only a 18\% contribution from $\tilde{A}\tilde{A}\to ZZ$ to the total annihilation, it is the channel that dominates the photon spectrum. The channels 
$\tilde{S}^\dagger\tilde{A}\to h\tilde{S}$(20\%)   and $\tilde{S}^\dagger\tilde{A}\to Z\tilde{S}$ (6\%) give sub-dominant contributions. Note that both of these benchmarks are within the reach of CTA.

%$${\SI[tight-spacing=true]{3.68e-4}}$$
\setlength{\tabcolsep}{4pt}
\begin{table}[!htb]
\begin{center}
\begin{tabular}{|cccccccccc|}
\hline
& $M_S$ & $M_A$  & $M_H$ & $M_{H^+}$ &$\lambda_{S1}$ &$\lambda_{S2}$ &$\lambda_{3}$ & $\lambda_{S12}$ & $\lambda_{S21}$\\
A&  244.3&411.9 & 449.8 & 418.5 & $\phantom{-}3.68\!\times\!\! 10^{-4}$   & 0.470&  $\phantom{-}9.07\!\times\!\! 10^{-5}$ &\phantom{-}1.86 &\phantom{-}0.98\\
B & 328.1&544.4&758.4&551.1&$-1.29\!\times\!\! 10^{-2}$& $ -1.02\!\times\!\! 10^{-3}$ & $ -3.18\!\times\! \!10^{-2}$ & $ -0.0103$ & -1.16\\
     \hline
     \end{tabular}
     \caption{Benchmark points}
     \label{tab:bench}
     \end{center}
\end{table}

We include all channels in the analysis that determines the reach of CTA. For this, we follow the analysis in ~\cite{CTA:2020qlo}. For an arbitrary spectrum, a table containing  the contribution to the total likelihood from each energy bin is provided. We compute the total spectrum with  micrOMEGAs for each point in our scan, and compare with the  total annihilation cross-section that can be probed for a given DM mass. Doing so, we find that CTA will significantly extend the reach of FermiLAT and AMS in probing the model, see Fig.~\ref{fig:fermi_ams}(left)  which shows  the exclusion in the $\langle \sigma_v\rangle_{VV} - M_A$ plane. Note that all points that are currently excluded by AMS-02 for the two hypothesis we considered for the DM density are also within the reach of CTA. The exclusion does not come only from the WW/ZZ channels, as we have checked that all the points that are within reach of CTA have a total $\langle\sigma_v\rangle > 3\times 10^{-26}{\rm cm}^3/{\rm s}$. For example the points with the lowest cross section in Fig.~\ref{fig:fermi_ams} are in fact dominated  by one of the semi-annihilation channels. 
 Note  that in some cases some large values of $\langle \sigma_v\rangle$ remain  beyond reach. This occurs either because invisible channels give a large contribution or because dominant channels lead to soft photons as illustrated for the $\tilde{H}^+W^-$ final state in the benchmarks above. Moreover note that  CTA is not sensitive to thermal cross sections for DM masses below about 200 GeV when it annihilates into W pairs~\cite{CTA:2020qlo}, hence many points  below 200 GeV cannot be probed.

 \begin{figure}[hbt]
	\includegraphics[scale=0.45]{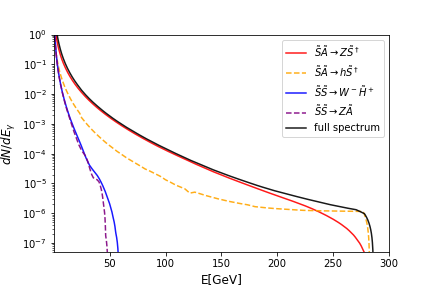}
	\includegraphics[scale=0.45]{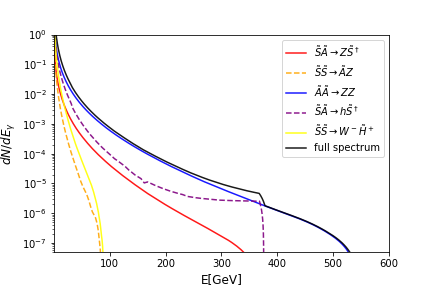}	
	\caption{ Photon spectrum for the two benchmarks A and B of Table~\ref{tab:bench}. }   
	\label{fig:spectrum} 
\end{figure}

 \subsubsection{Colliders, Direct and Indirect detection}
 \label{sec:general_coll}
 
 Before doing a comparison of the reach in future direct and indirect searches, we briefly discuss collider constraints. From the general scan we considered only the points that were not excluded by other constraints (theoretical, relic density and  Xenon-1T). To evaluate the current and future probes of the model through monojet, we use the analysis of ~\cite{Belyaev:2018ext} performed in the IDM. 
 We computed separately the cross-section for monojet production corresponding to  $pp\to \tilde{A}\tilde{A}j$ as well as $pp\to \tilde{S}\tilde{S}j$.  The former is driven mainly  by $\lambda_{Ah}$ while the latter depends on $\lambda_{S1}$.   Since $\sigma(pp\to \tilde{A}\tilde{A}j)= 1/2 \sigma(pp\to \tilde{S}\tilde{S}j)$ for $\lambda_{Ah}=\lambda_{S1}$, we simply translated the limit on $\lambda_{Ah}$ in a limit on $\lambda_{S1}$. These limits together with our parameters points are displayed  in the plane $\lambda_{Ah}-M_A$ and $\lambda_{S1}-M_A$ respectively. For $M_{DM}>80$GeV, even the high luminosity LHC can only probe  couplings ${\cal O}(1)$ whereas, as discussed above in our model the couplings are generally between $10^{-3}-10^{-1}$. Couplings $\approx 10^{-2}$ can be probed if $M_{DM} < 70$~GeV but we find only a few points where $M_A$ lies in that mass range. Indeed the lighter doublet is constrained both by LEP, by charged Higgs searches  as well as various theoretical constraints. We also note that the singlet contribution to the monojet is always far from the experimental limit especially since we find few points in the low mass region, the approximation of computing separately the doublet and singlet contribution is therefore valid. Thus we conclude that the monojet search cannot probe the Z4IDSM once other theoretical or dark matter constraints are taken into account.  
 
  \begin{figure}[h]
	\includegraphics[scale=0.45]{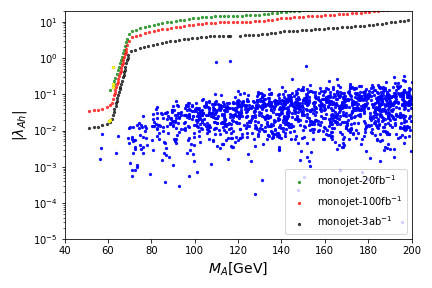}
	\includegraphics[scale=0.45]{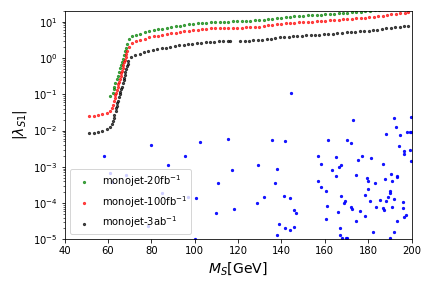}
	\centering
	\caption{ Monojet limit at LHC for ${\cal L}=20,100,3000 {\rm fb}^{-1}$ and allowed points  for the Z4IDSM model (blue) in the $|\lambda_{Ah}| - M_A$ and $|\lambda_{S1}| - M_S$ plane.}   
	\label{fig:monojet} 
\end{figure}

 Finally we perform a comparison of the direct and indirect detection reach.  Figure 14 shows  the reach for XENON-nT, CTA and DARWIN   in the $\langle \sigma_v \rangle_{VV}$ - $M_A$ plane  for points that pass all constraints including those from FermiLAT and AMS02. The same points are plotted in the $\sigma_{Ap} \xi_2$ - $M_A$ plane.  A large fraction of the points are within reach of both CTA and XENON-nT and/or DARWIN, this large overlap between the DD and ID probes opens the possibility of cross-checking a DM signal, nevertheless the parameter space will not be fully covered.
XENON-nT   will probe basically all scenarios where ${\lambda}_{Ah} \geq 0.1$ or $\lambda_{S1} >0.071$, while DARWIN will improve these by roughly a factor 2, in addition the bound on  $\lambda_{S1}$ is stronger by nearly an order of magnitude when the DM masses are near the electroweak scale.  CTA will extend the reach to cases where the  SI signal for both DM  is weak, notably for the doublet near the TeV scale. CTA  however does not cover completely the region with  lighter masses of the doublet (below 300 GeV), both because the signal is suppressed by the small value of $\xi_2$ and because the sensitivity is not as good. In fact there are cases where both $M_S$ and $M_A$ are below 300 GeV that cannot be probed, typically because the couplings  ${\lambda}_{Ah}$ and $\lambda_{S1}$ are small. For these points either semi-annihilation or DM conversion plays an important role.
We found that scenarios that are out of reach cover the full range of DM masses and in general correspond to a subdominant doublet DM with 
$\Omega_2 h^2$  between .001 and .02.
 
  \begin{figure}[h]
	\includegraphics[scale=0.5]{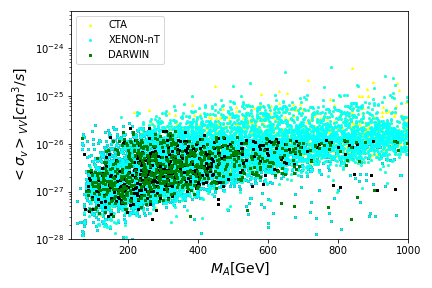}	
	\includegraphics[scale=0.5]{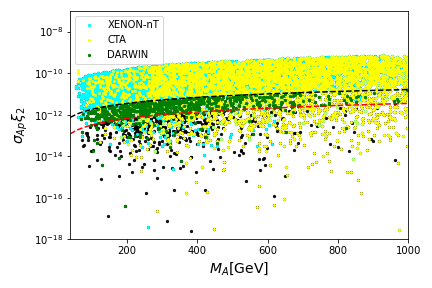}	
	\caption{Future reach of direct detection with XENON-nT (cyan) and  DARWIN (green) and indirect detection with photons (CTA) in the $\langle\sigma_v\rangle_{VV}-M_A$ (left)  and  $\sigma_{Ap}\xi_2-M_A$ (right) planes. Points in black are beyond the reach of these detectors. In the right panel the projected exclusions for XENON-nT (black-dash) and DARWIN (red-dash) in the case of one component DM are also shown.}   
	\label{fig:futurereach} 
\end{figure}

\subsection{The nearly degenerate doublet}
\label{sec:ND}

In order to investigate more closely the case where the doublet is nearly degenerate, we performed an independent scan. A compressed spectrum opens up the possibility of 3 DM candidates, indeed for very small mass splittings between the two neutral component of the doublet, both can be stable and contribute to the relic density. 
To investigate this possibility more precisely, we assume that the two neutral components of the doublets are nearly degenerate,   and we treat the mass splitting of the charged component as a free parameter in the range 
\begin{equation}
\Delta^+ = 1-500 {\rm MeV}
\end{equation}
Note that we set $\Delta^0=201{\rm keV}$  corresponding to the minimum value that prevents  inelastic scattering on nucleons, a process which is strongly constrained since it  depends on the gauge coupling $\tilde{A}\tilde{H}Z$~\cite{LopezHonorez:2006gr}.  In this case $M_H$ will be nearly stable and can be treated as a third DM component when computing the relic density.

The remaining masses  are varied randomly in the range $40 {\rm GeV}<M_S< 1.5 {\rm TeV}$ and $70 {\rm GeV}<M_A< 1 {\rm TeV} $ while we use the range defined in Table~\ref{tab:range} for the couplings.
Note that with this choice of mass splitting amongst the neutral components of the doublet, the coupling $\lambda_5 \approx 10^{-7}- 10^{-6}$, thus the coupling of the scalar and the pseudoscalar to the Higgs are nearly equal $\lambda_{Ah}\approx \lambda_{Hh}$. This entails that the contribution of $\tilde{H}$ and $\tilde{A}$ to the relic density are nearly equal.

\subsubsection{Collider constraints}

\begin{figure}[h]
	\includegraphics[scale=0.5]{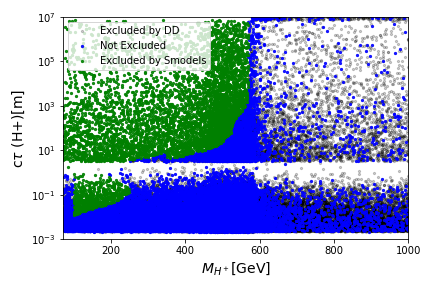}
	  \includegraphics[scale=0.5]{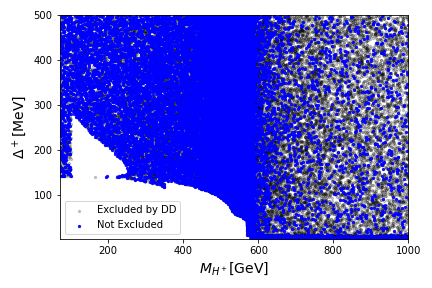}
	\centering
	\caption{Left: LHC exclusion from disappearing tracks and HSCP (green) in the $c\tau - M_{H^+}$ plane after applying relic density and direct detection constraints from XENON1T (black), allowed points are in blue. Right: Points allowed after LHC constraints in the $\Delta^+ - M_{H^+}$ plane. }
		\label{fig:ctau_ND} 
\end{figure}

As mentioned in section~\ref{sec:LHC}, the new charged scalar can be long-lived and lead to a signature in disappearing track or HSCP. Current LHC limits exclude long-lived $\tilde{H}^+$ with masses below 550 GeV, see Fig.~\ref{fig:ctau_ND}. The mass limit weakens for shorter lifetime, for example when $c\tau \approx 10m$ the mass limit drops to 500~GeV and 
when $c\tau \approx 0.5m$ only charged Higgs lighter than 200 GeV are excluded.
The disappearing track search targets lower lifetimes $c\tau< 0.3m$ and can reach only up to $M_{H^+} \approx 250 GeV.$  Note that the disappearing track search at LHC does not cover the region $M_{H^+} <100$GeV. The typical mass split that is constrained by HSCP is below 150MeV while the disappearing track can apply for larger $\Delta^+$ up to 280MeV,
see Fig.~\ref{fig:ctau_ND} (right).

 Note that, as we will discus in the next section, we found a   more stringent limit on the coupling $\lambda_{Ah}$ as compared with the general case described in section~\ref{sec:general_coll} we do not expect significant constraints from monojet searches, see also~\cite{Belyaev:2018ext}.

\subsubsection{DM observables : relic density and direct detection}

After imposing collider constraints we explore DM observables. 
To solve for the DM relic density, we use Eq.~\ref{eq:Y3} that include a third DM component, $\tilde H$. Because we have fixed $\Delta^0$ to be very small, $\tilde H$ has not decayed at the time of DM formation. We compute its relic density and find that it is approximately the same as that of $\tilde A$. 
In the following results we understand $\Omega_2 =\Omega_H +\Omega_A$ to be the total contribution of the doublet to the relic density. Note however  that this third component will decay into $\tilde{A}$ and will not be present in the Universe today. To check this we have solved the abundance equations, Eq.~\ref{eq:Y3} until $T=10^{-13}$GeV and found $Y_{\tilde{H}}$ to be negligible. 

\begin{figure}[h]
	\includegraphics[scale=0.5]{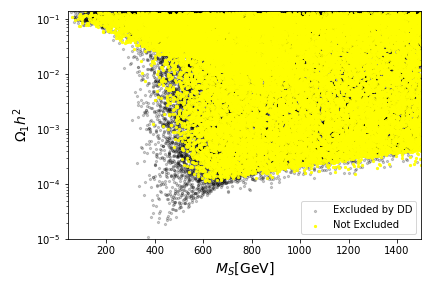}
	\includegraphics[scale=0.5]{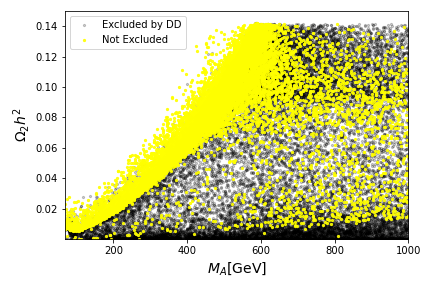}
	\centering
	\caption{$\Omega_1h^2$ (left) and $\Omega_2h^2$ (right) as a function of the corresponding  DM mass, for all points satisfying theoretical, collider and the total relic density constraints (grey) as well as DD constraint (yellow).}
	%{\bf for Omega1 increase the upper limit os there is a bit of free space at the top of the plot}  }   
	\label{fig:omega12_ND} 
\end{figure}

As known from the IDM, when the doublet is compressed, its relic density can be in agreement with PLANCK for masses above roughly 500-600 GeV. This is the main feature that we recover in the Z4IDSM compressed model, Fig.~\ref{fig:omega12_ND} right.  For lighter masses, the doublet is a subdominant DM component and $\Omega_2$ increases with $m_A$ up to about 500 - 600 GeV. This means that it is much easier to have a dominant doublet DM than in the general scan. For heavier masses of the doublet, we would expect the doublet to be overabundant, however the effect of DM conversion and/or semi-annihilation can help reduce its relic density thus we find allowed scenarios for $M_A$ covering the full range of masses probed. At the same time $\Omega_1$ is likely to be subdominant especially at heavier masses.

\begin{figure}[htb]
	\includegraphics[scale=0.5]{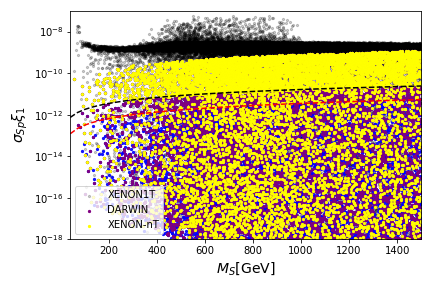}
	\includegraphics[scale=0.5]{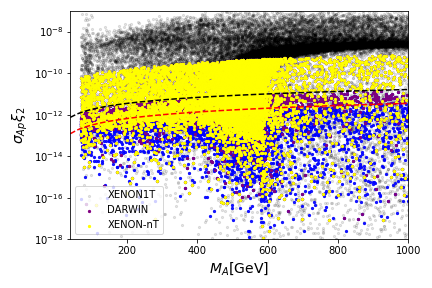}
	\centering
	\caption{Spin-independent DM proton scattering cross section normalized with  the fraction of DM component,  as a function of the corresponding DM masses, $\sigma_{Sp} \xi_1 -M_S$ (left) and $\sigma_{Ap} \xi_2 -M_A$ (right). The black (red) line shows the projected reach of XENON-nT (DARWIN), points in yellow (purple) are within the reach of XENON-nT (DARWIN), and blue points are beyond the reach of DARWIN. }
	\label{fig:ND-Darwin} 
\end{figure}

\begin{figure}[h]
 \includegraphics[scale=0.5]{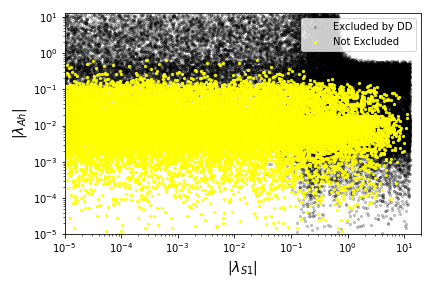}	
       \includegraphics[scale=0.5]{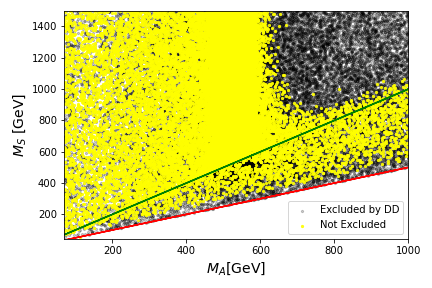}	
	\centering
	\caption{Left: Allowed points (yellow) in the $|\lambda_{Ah}| - |\lambda_{S1}|$ plane(left) and in the $M_S -M_A$ plane (right). Also shown are points  that are excluded by XENON-1T (black). }   
	\label{fig:ND_MSMA} 
\end{figure}

\begin{figure}
	\includegraphics[scale=0.5]{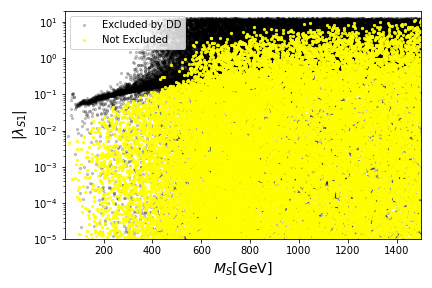}
	\includegraphics[scale=0.5]{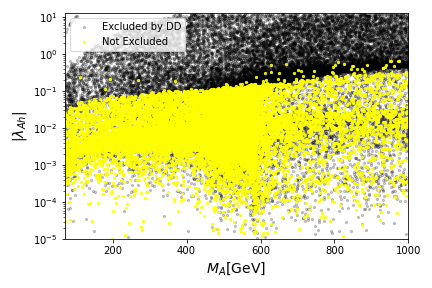}
	\centering
	\caption{ All points satisfying theoretical, collider and the total relic density constraints (grey) as well as DD constraint (yellow) in the $\lambda_{S1}-M_S$ plane (left) and  $\lambda_{Ah}-M_A$ plane (right) . }   
	\label{fig:lambdaS1_ND} 
\end{figure}

The current constraints and future reach of DD experiments are displayed  in Fig.~\ref{fig:ND-Darwin}. We see in particular that a much larger fraction of points can be probed by XENON-nT through the singlet component as compared with the general scan. Those are the points in yellow that lie below the XENON-nT projected limit on  $\sigma_{Ap} \xi_2$ in Fig.~\ref{fig:ND-Darwin} right. This is tied to the  main difference with the general scan  which is found in the range of values for the couplings $\lambda_{S1}$ and $\lambda_{Ah}$ which are allowed after imposing the DD constraints from XENON-1T, Fig.~\ref{fig:ND_MSMA}(left).
With a compressed spectrum it is possible to have $\lambda_{S1}>1$  while escaping  DD constraints, see Fig.~\ref{fig:lambdaS1_ND} (left). The  reason for this is that since  $\tilde{S}$ can be only a sub-dominant DM component, its DD signal, which is proportional to the  DM fraction $\xi_1$, can be suppressed despite a large value of $\lambda_{S1}$. Indeed large values of $\lambda_{S1}$ are found for   $\Omega_S <<0.1$. This is only possible for  $M_S>500$GeV since it requires the doublet to be the dominant DM component. 
For the same reason,  because the doublet contributes significantly to the total DM component, it cannot evade so easily DD constraints, thus the upper limit on the coupling $\lambda_{Ah}<0.2$  that is set by the XENON-1T constraint is more stringent than for the general scan~\ref{fig:lambdaS1_ND}. Note that  DD constraints are particularly strong when $M_A$ is in the range 400-600 GeV and the doublet is the dominant DM component, as seen by the high density of points  in Fig.~\ref{fig:lambdaS1_ND} (right) in that region. 
As opposed to the general scan it is not possible to have both $\tilde{S}$ and $\tilde{A}$ very heavy, Fig.~\ref{fig:ND_MSMA}(right).In this region, despite the contributions  of semi-annihilation and DM conversion, large values for $\lambda_{S1}$ and $\lambda_{Ah}$ are required to ensure that both $\Omega_i$ are small enough, such couplings are constrained by XENON1T. 

\begin{figure}[h]
       \includegraphics[scale=0.5]{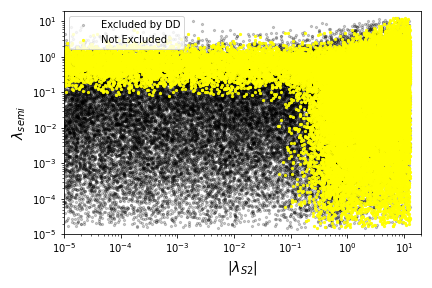}	
             \includegraphics[scale=0.5]{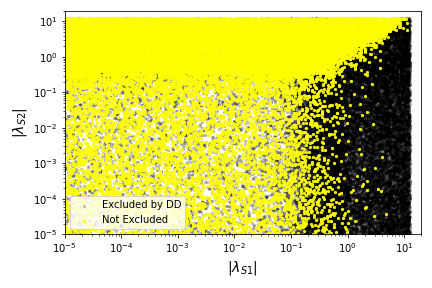}
	\centering
	\caption{Allowed points in the $\lambda_{semi}$ - $|\lambda_{S2}|$ plane (left) and $|\lambda_{S2}|$ - $|\lambda_{S1}|$ plane (right). }   
	\label{fig:semi_conv_ND} 
\end{figure}

With the compressed spectrum we also find that as in the general scan some amount of semi-annnihilation or DM conversion  are necessary to evade DD constraints, indeed  in Fig.~\ref{fig:semi_conv_ND} (left) we clearly see that DD imposes a lower bound on  $\lambda_{semi} , \lambda_{S2} \approx 0.1$. Moreover we find that a much larger  fraction of the allowed points are associated with  $\lambda_{S2}>1$ as compared to the general case. The reason for this was discussed in section~\ref{sec:relic},  the assisted FO mechanism in presence of a large cross-section $\sigma^{1122}$ will reduce the abundance of both DM components, but more significantly that of the heavier component, which is generally the singlet. Therefore for large $\lambda_{S2}$ the abundance of the singlet will be very small, thus the doublet has to account for all of the DM. This is hard to realise 
 when the spectrum is not compressed since typically $\Omega_2<<0.1$ even without the additional contribution from the DM conversion term. When the singlet is the lighter component, we also find allowed points with $\lambda_{S2}>1$, those also feature a large semi-annihilation coupling $\lambda_{semi}>1$, in this case processes such as $\tilde{S}\tilde{S}\to \tilde{A}Z, \tilde{A}h$ contribute significantly to bring $\Omega_1 h^2$ below the PLANCK limit while DM conversion process suppress $\Omega_2 h^2$.

While both the singlet and doublets DM  can be within reach of future DD detectors, a large fraction of the model parameters remains beyond the reach of XENON-nT and even DARWIN.  In part this is related to the fact that, as mentioned above,  the coupling of the DM to the Higgs is suppressed, with much lower values of $\lambda_{Ah}$ favoured.

\subsubsection{Indirect Detection}

The largest annihilation cross-sections  for the doublet are into  $WW/ZZ$ pairs, they can reach  ($\approx 10^{-25} {\rm cm}^3/{\rm sec}$) when $M_A\approx 400$~GeV, in this  case  the doublet form the dominant DM component.   Typically the WW channel is slightly larger than the $ZZ$ channel since all doublet components  are nearly degenerate. As in the non-compressed case other SM final states are subdominant. The only other important annihilation channel for the doublets are into pairs of singlets, although this cross-section can exceed $10^{-25}  {\rm cm}^3/{\rm sec}$ it does not contribute to the signal. 

The pair annihilation channels $\tilde S \tilde S \to SM,SM $ reach at most  $7\times 10^{-27} {\rm cm}^3/{\rm sec}$ in the WW channel. As in the general case other final states that can exceed $10^{-26}{\rm cm}^3/{\rm sec}$ include most of the semi-annihilation $\tilde S {\tilde S}^\dagger \to \tilde{H}^\pm W^\mp,\tilde{A}h,\tilde{A}Z, \tilde{H}h,\tilde{H}Z$ or DM conversion processes $\tilde S {\tilde S}^\dagger \to \tilde{H}^+\tilde{H}^-$. Moreover semi-annihilation channels such as  $\tilde{S}^\dagger \tilde A\to h\tilde S,Z\tilde S$ can be dominant, both  can reach approximately $10^{-25} {\rm cm}^3/{\rm sec}$ for $M_A>600$GeV .

\begin{figure}[h]
	\includegraphics[scale=0.5]{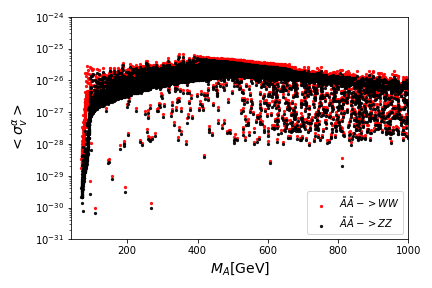}
	\includegraphics[scale=0.5]{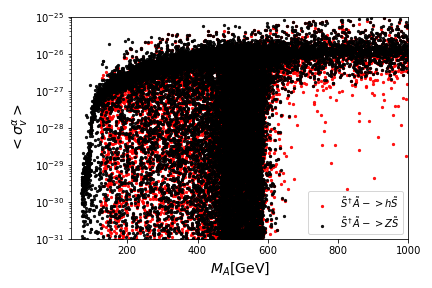}
	\includegraphics[scale=0.5]{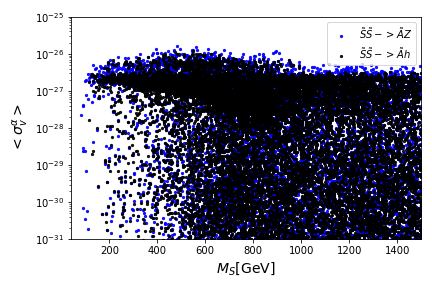}	
	\includegraphics[scale=0.5]{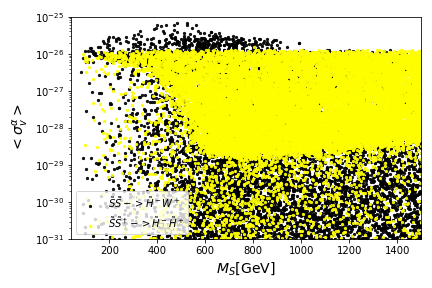}	
	\centering
	\caption{Cross-sections for the main  DM annihilation and semi-annihilation channels indicated in each panel. 
	} 
	\label{fig:ND_ID_channels} 
\end{figure}

\begin{figure}[hbt]
	\includegraphics[scale=0.5]{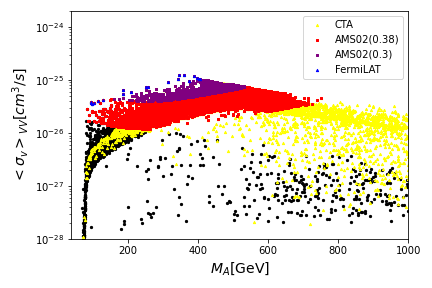}
	\includegraphics[scale=0.5]{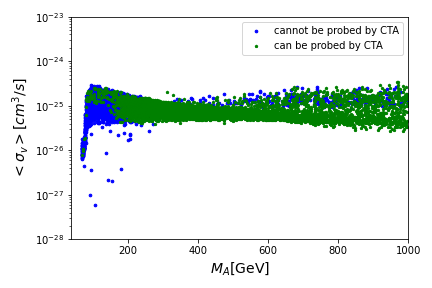}
	\centering
	\caption{ Left panel : Constraints on  $\langle \sigma_v\rangle_{VV}$ from FermiLAT (blue)  and AMS02 assuming an NFW profile with $\rho_{DM}=0.3$(purple) or a generalised NFW profile with  $\rho_{DM}=0.38$(red). The reach of CTA is also displayed (yellow). Black points are beyond the reach of these three indirect detection searches. Right panel :   Total $\langle \sigma_v\rangle$  corresponding to the points that are within (yellow) or beyond (blue) reach of CTA. }
	\label{fig:ND_Fermi_AMS} 
\end{figure}

We again use a conservative approach and include only the WW and ZZ channels to derive the FermiLAT and AMS02 constraints. 
 However, for detectors with increased sensitivity, such as CTA, it becomes more important to exploit all channels as was illustrated in the general scan. Recall that in the WW channel CTA is expected to probe $\langle \sigma_v\rangle ~\approx 2(1) \times10^{-26}{\rm cm}^3/{\rm sec}$ for $M_{DM}=0.2(1) {\rm TeV}$.
 The constraints are displayed in terms of $\langle \sigma_v\rangle_{VV}$ in Fig.~\ref{fig:ND_Fermi_AMS}(left). FermiLAT only constrain a few scenarios corresponding to the largest annihilation cross-section into WW while AMS02 provides stronger constraints for $M_A$ up to 700 GeV  depending  on the assumption on the DM profile and the local density. CTA will also probe the scenarios with large cross-section in WW and will extend the reach to the highest mass used in the scan (1TeV). Moreover because of the important contribution from semi-annihilation channels, CTA  will also probe many scenarios where the WW/ZZ channels are suppressed. In Fig.~\ref{fig:ND_Fermi_AMS} (right), one can see that 
the total $\langle \sigma_v \rangle$ is always above $3\times 10^{-26}{\rm cm}^3/{\rm sec}$ except when $M_A\approx M_h/2$,  yet several scenarios remain out of reach of CTA,  in particular when $M_A<200$GeV. This is  explained by the limited sensitivity of the experiment in that region and by the fact that   the contribution from the WW/ZZ channels which give generally the  main signature  can be quite suppressed. In addition, for a large number of points DM conversion processes $\tilde S\tilde S^\dagger \leftrightarrow \tilde A\tilde A$ dominate DM annihilation, leaving no signature for indirect detection. This can occur for any value of $M_A$. 
 
\subsubsection{Complementarity Colliders, Direct and Indirect detection }

\begin{figure}[hbt]
	\includegraphics[scale=0.5]{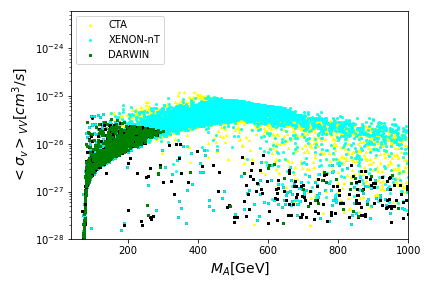}
	\includegraphics[scale=0.5]{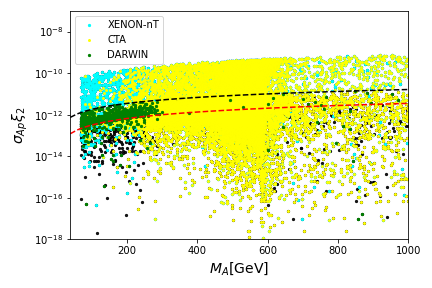}	
	\centering
	\caption{Reach of XENON-nT (blue), DARWIN (green), CTA (yellow) for $\langle \sigma_v\rangle_{VV}$ (left) and $\sigma_{\tilde Ap} \xi_2$  (right) {\it vs}  $M_A$. Only  points that are allowed by current constraints are included, for DARWIN  only points that are beyond the reach of XENON-nT are considered. Black points are beyond the reach of all detectors.}
	\label{fig:ND-ID_DD} 
\end{figure}

To highlight the complementarity between the indirect and direct detection searches, 
we compare the future reach  of CTA, XENON-nT and DARWIN in  Fig.~\ref{fig:ND-ID_DD} expressed in terms of  $\langle \sigma_v\rangle_{VV}$ and $\sigma_{Ap} \xi_2$ as a function of $M_A$.   As in the general case, most of the parameter space can be probed. Clearly the low mass region ($M_A<200 {\rm GeV}$) will be better probe by DD detectors, since  in this region CTA has a reduced sensitivity. From Fig.~\ref{fig:ND-ID_DD}, it might seem that there is a large overlap for the points that are within reach of CTA and those that are within  reach of XENON-nT or DARWIN, however the two types of searches probe different scenarios. DD can probe scenarios with large values of $\lambda_{S1}$ or $\lambda_{Ah}(\lambda_{Hh})$ as seen in Fig.~\ref{fig:lambdaS1_ND}, on the other hand CTA can probe points where $\tilde{A}\tilde{A}\to WW$ is large or points with large values for $\lambda_{semi}$ and/or $\lambda_{S2}$ by taking advantage of semi-annihilation channels.  The former   requires that the doublet component, $\xi_2$  is not too suppressed while the latter can depend on both $\xi_1$ and $\xi_2$ according to the dominant process.  

A few benchmarks that illustrate this complementarity between ID and DD  are shown in Table~\ref{tab:bench_compressed} for DM masses of a few hundred GeV's. All points have $\langle \sigma_v\rangle > 4 \times 10^{-26}{\rm cm}^3/{\rm sec}$, however only BP1 and BP3 are within reach of CTA, due to the large contribution from $\tilde{A}\tilde{A}\to WW$ with $\xi_2 \approx 0.5$ in both cases.
Note however that there are points with $\lambda_{S2}>1$ that remain out of reach of CTA - we found that all these points fell in the region where  $M_S<M_A$ such that  $\tilde{A}\tilde{A}\to \tilde{S}\tilde{S}^\dagger$ is dominant, leaving no signal.  
BP1 is an example of a point with large $\lambda_{S2}$ and large contribution from $\tilde{A}\tilde{A}\to \tilde{S}\tilde{S}^\dagger$ that has nevertheless a large enough contribution from $\tilde{A}\tilde{A}\to W^+W^-$.
This point escapes future DD searches due to the small values of $\lambda_3\approx \lambda_{Ah}$ and $\lambda_{S1}$. BP2 on the other hand, with similar masses and values of $\lambda_{S2}$ escapes CTA's reach, due to the suppressed contribution of the doublet to dark matter. Moreover since 
$\lambda_3\approx \lambda_{Ah} \approx 10^{-2}$,  the doublet component can be probed by DARWIN.  BP3 features a similar value of $\lambda_{Ah}$ and also has $\lambda_{S1}$  large enough such  that both the singlet  and doublet component are within DARWIN's reach. Finally BP4 is an example of a point that falls beyond the sensitivity of future detectors. Here $\lambda_3$ and $\lambda_{S1}$ are small, thus leading to weak DD signals while semi-annihilation processes driven by $\lambda_{S12}$ ensures that the relic density constraint is satisfied. This point escapes CTA due to the somewhat suppressed $\langle \sigma_v\rangle_{VV}$ and the fact that the mass of the  doublet  200GeV falls in the region where the sensitivity of CTA is not maximal.

\begin{table}[!hbt]
\begin{center}
\begin{tabular}{|l|cccc|}
\hline
 &BP1&BP2&BP3&BP4\\\hline
 $M_A$ &313.594&327.952&360.31&202.74\\
 $M_{H^+}$&313.727&328.265&360.81&203.01\\
  $M_S$&313.1&317.98&848.9&287.8\\
  $\lambda_2$&0.185&$1.7\times 10^{-2}$&$7.5\times 10^{-3}$&$7.8\times 10^{-2}$\\
  $\lambda_3$&$1.03\times 10^{-5}$&$4.05\times 10^{-2}$&$2.5\times 10^{-2}$&$2.0\times 10^{-4}$\\
  $\lambda_{S1}$&$2.98\times 10^{-3}$&$1.77\times 10^{-4}$&$1.32\times 10^{-2}$&$2.15\times 10^{-3}$\\
    $\lambda_{S2}$&3.94&4.01&$6.17\times 10^{-1}$&$1.75\times 10^{-3}$\\
      $\lambda_{S12}$&$1.5\times 10^{-2}$&$3.11\times 10^{-3}$&$1.75\times 10^{-4}$&$1.46\times 10^{-1}$\\
        $\lambda_{S21}$&$4.08\times 10^{-2}$&$2.38\times 10^{-1}$&$7.3\times 10^{-3}$&$3.5\times 10^{-3}$\\
            $\Omega_1 h^2$&0.0641&0.112&0.0491& 0.105\\
            $\Omega_2 h^2$&0.0427&0.0146&0.0592&  0.0114\\
            $\sigma^{SI}_{\tilde{S} p}$ (pb)&$8.0\times 10^{-13}$&$2.74\times 10^{-15}$&$2.2\times 10^{-12}$&$3.9\times 10^{-13}$\\
        $\sigma^{SI}_{\tilde{A} p}$ (pb)&$6.8\times 10^{-13}$&$9.36\times 10^{-11}$&$1.2\times 10^{-11}$&$3.9\times 10^{-12}$\\
        DARWIN & $\times$& \checkmark &\checkmark &$\times$\\
             $\langle \sigma_v \rangle_{VV} \xi_2^2$ (cm$^3$/s)&$2.95\times 10^{-26}$  &$2.37\times 10^{-27}$  &$4.27\times 10^{-26}$ &$1.67\times 10^{-27}$\\
               $\langle \sigma_v \rangle$ (cm$^3$/s)&$2.0\times 10^{-25}$&$9.39\times 10^{-26}$&$8.25\times 10^{-26}$&$1.17\times 10^{-25}$\\
               CTA & \checkmark & $\times$ &\checkmark &$\times$\\
               $c\tau$(m)&4.2   & $9.9\times 10^{-3}$ & $2.3\times 10^{-3}$  & $7.8\times 10^{-3}$ \\
     \hline
     \end{tabular}
     \caption{Benchmarks}
     \label{tab:bench_compressed}
     \end{center}
\end{table}

\begin{figure}[hbt]
	\includegraphics[scale=0.5]{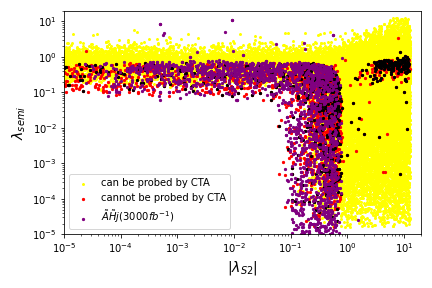}	
		\includegraphics[scale=0.5]{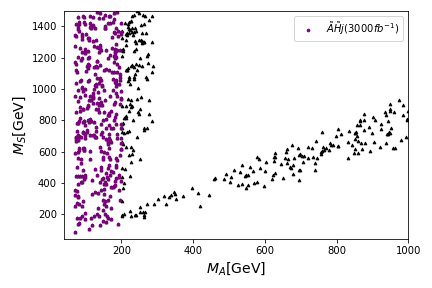}
	\centering
	\caption{Left: Points that can (yellow) or cannot (red) be probed by CTA, in purple scenarios that are within reach of HL-LHC and in black points that are beyond the reach of CTA,DARWIN and HL-LHC.  
	Right: Masses of singlet and doublet dark matter candidates for scenarios that remain beyond the reach of future DD and ID detectors but are within reach of HL-LHC (purple). }
	\label{fig:ND-beyond} 
\end{figure}

The scenarios that remain out of reach of the future detectors considered are concentrated in two regions:   $M_A<300$GeV and  $M_A/2<M_S<M_A$, see Fig.~\ref{fig:ND-beyond} right. 
For these scenarios, the singlet is the dominant DM component, as was the case for BP4. Because of the direct detection constraint, 
these scenarios have a small value for both $\lambda_{S1}$ and $\lambda_{Ah}$,  which are restricted to be below $0.01-0.03$,  with the stricter bound for lighter $M_A$. Thus an important contribution from semi-annihilation and DM conversion is needed to satisfy the relic density constraint. Indeed we find that a significant fraction of the points that are out of reach correspond to   $\lambda_{S2}>1$, they fall in the region where  $M_S<M_A$ and as mentionned above are invisible because of the large contribution from  $\tilde{A}\tilde{A}\to \tilde{S}\tilde{S}^\dagger$.  In the region $M_A\approx 200$GeV, $\langle \sigma_ v\rangle$ can exceed $10^{-26} {\rm cm}^3/{\rm s}$, however the small value of $\Omega_2$ leads to a weaker signal that is not always compensate by a large contribution from semi-annihilation channels. For example for BP4,  $\tilde{S}\tilde{S}\to V \tilde{D}$ where $\tilde{D}$ stands for any component of the doublet has a cross-section
$\langle \sigma_v \rangle_{SD}  = 7.6\times 10^{-27} {\rm cm}^3/{\rm s}$. 

Concerning the potential of the high luminosity LHC (HL-LHC) to probe the nearly degenerate scenario, we mention that an analysis of the $pp\to \tilde{A}\tilde{H}j$ leading to a monojet signature in the case of a compressed spectrum of the doublet  ~\cite{Belyaev:2018ext}, showed that the HL-LHC with a luminosity of 3ab$^{-1}$ could constrain doublet DM masses up to 200 GeV independently of the value of $\lambda_{Ah}$.  Such region could also be probed at a future 100 TeV collider~\cite{Blinov:2015qva}. This coverage is complementary to the one from astrophysical searches. In particular it will allow to cover the low mass region in Fig.~\ref{fig:ND-ID_DD} that will be  mainly probed by a detector like DARWIN. Moreover the HL-LHC would cover most of the unprobed region where the singlet is the heavier DM, see Fig.~\ref{fig:ND-beyond}. Finally note that searches for disappearing tracks at upcoming runs at LHC could allow to probe some of the points that escape astrophysical and collider searches. For example, BP4 has  $c\tau=0.78{\rm cm}$  and is just beyond the reach of current searches. Note however that $c\tau$ depends critically on the mass splitting which we chose here to be below 500MeV.

\section{Discussion}

Enlarging the dark matter sector of the IDM and Singlet model  to have two (or three) DM candidates opens up significantly the possibility for DM compatible with stringent relic density constraints and direct detection constraints. Because of the presence of semi-annihilation and/or DM conversion  processes the relic density constraint can be satisfied with reduced couplings of each DM component to the Higgs, thus escaping the direct detection constraints. We have investigated the parameter space of the model that satisfies all current constraints
and found that generally the singlet is the dominant DM component. We also found that the mass range up to the TeV scale can be satisfied, although strong collider constraints preclude DM much below the electroweak scale. In fact after taking into account direct detection and relic constraint the invisible width of the Higgs is sufficiently suppressed that it does not add significant constraint. We also highlighted the complementarity of direct and indirect searches and stressed the importance of including semi annihilation processes such as $\tilde{S}^\dagger \tilde{A}\to Z\tilde{S}, h\tilde{S}$ that can contribute significantly to the photon spectrum in indirect detection. After taking into account cosmological and astrophysical constraints and requiring that both DM components explain totally the relic density, we found that the conventional DM signatures at colliders, such as monojet  did not help in further probing the model. 
While future astrophysics search will provide powerful probe of the model, some scenarios can still escape future searches by CTA and DARWIN, typically those scenarios have a doublet below 400 GeV, feature a subdominant doublet component and have small couplings to the Higgs.

Since a compressed doublet can lead to a quite different phenomenology we have investigated separately the case of a very degenerate doublet spectrum, while allowing the singlet to be any mass. In this case the doublet can be the dominant dark matter component. In fact for the very compressed spectrum that we considered the two neutral components of the doublet contribute equally to the relic density, for this we solved the generalized Boltzmann equation for 3 DM, although we present the results summing over the contributions of the two doublet component. Because of the near mass degeneracy, the charged scalar can lead to signatures in both disappearing tracks and searches for heavy stable charged particles at the LHC. These signatures are however very much dependent on the exact amount of compression with the charged and neutral Higgses. In this case we also illustrate the complementarity of current and future direct and indirect detection searches and show that while most of the parameter space could be probed, some scenarios escape detection. As in the general case they correspond to cases where the couplings of DM to the Higgs is suppressed and there is a significant amount of semi annihilation or DM conversion. Because semi-annihilation often leads to much softer photon spectrum in indirect detection, improving the sensitivity at low energies would be required to probe all scenarios. In particular we found that when DM conversion was important a pair of doublet could annihilate into a pair of singlets if kinematically accessible rendering both DM invisible.

In this paper we considered only the case where the singlet and the doublet are WIMPs, in general  in this model the couplings of the singlet can be so weak that it  is feebly interacting and is produced via freeze-in in the early Universe. A detailed analysis of this case is left for a separate analysis.

\section{Acknowledgements}

We thank C. Eckner for helpful discussions on CTA and the use of his code for reading the spectrum tables. We  thank Shankha Banerjee for checking the one-loop contribution to the charged Higgs partial decay width. 
We also acknowledge useful discussions with Alexander Belyaev, Fawzi Boudjema, Bjorn Herrmann  and Sabine Kraml.
This work  was funded by RFBR and CNRS, project number 20-52-15005. The work of A. Pukhov was also supported in part by a grant AAP-USMB.

\bibliography{z4wimp}

\end{document}